\documentclass[12pt]{article}
\usepackage{amssymb}
\usepackage{epsfig}
\usepackage{array}
\usepackage{float}
\usepackage{amstext}
\usepackage{amsmath}
\usepackage{psfrag}
\usepackage{a4}
\usepackage{a4wide}
\usepackage{hyperref}
\usepackage{graphicx}
\usepackage{subcaption}
\usepackage{epsfig}
\usepackage{epstopdf}
\usepackage{xcolor,colortbl}

\usepackage[T1]{fontenc}

\DeclareSymbolFont{myletters}{OML}{ztmcm}{m}{it}
\DeclareMathSymbol{\uplambda}{\mathord}{myletters}{"15}
\DeclareMathSymbol{\upxi}{\mathord}{myletters}{"18}

\usepackage{amsfonts}

\usepackage{lmodern} 
\usepackage{amsmath}                                                    
\usepackage{amsthm}                                                     
\usepackage{amssymb}
\usepackage{multirow}
\numberwithin{equation}{section} 
\newcommand{\newc}{\newcommand}
\newc{\be}{\begin{equation}}
\newc{\ee}{\end{equation}}
\newc{\bg}{\begin{gathered}}
\newc{\eg}{\end{gathered}}
\newc{\tref}[1]{Table \ref{#1}}
\newc{\eref}[1]{Equation \eqref{#1}}
\newc{\su}[1]{$SU(#1)$}
\newc{\bm}[1]{\mathbf{#1}}
\newc{\fref}[1]{Figure \ref{#1}}

\newc{\ra}{\rightarrow}
\newc{\lra}{\leftrightarrow}
\newc{\ov}{\overline}
\newc{\ba}{\begin{eqnarray}}
\newc{\ea}{\end{eqnarray}}
\newc{\mf}{\mathsf}

\usepackage{hyperref}
\usepackage{ulem}
\DeclareMathOperator{\tr}{Tr}
\newcolumntype{g}{>{\columncolor{Gray}}c}
\newcolumntype{w}{>{\columncolor{Gray2}}c}
\definecolor{Gray}{gray}{0.95}
\definecolor{Gray2}{gray}{0.98}
\def\be{\begin{equation}}
\def\ee{\end{equation}}
\def\bea{\begin{eqnarray}}
\def\eea{\end{eqnarray}}

\begin{document}
	
	\begin{center}
		\baselineskip 20pt 
		{\Large\bf Observable $r$, Gravitino Dark Matter, and Non-thermal Leptogenesis in No-Scale Supergravity}
		
		\vspace{1cm}

		{\large 
			\textbf{Waqas Ahmed}$^{a}$ \footnote{E-mail: \texttt{\href{mailto:waqasmit@hbpu.edu.cn}{waqasmit@hbpu.edu.cn}}},
		\textbf{	Muhammad Moosa}$^{b}$ \footnote{E-mail: \texttt{\href{muhammad\_moosa@comsats.edu.pk}{muhammad\_moosa@comsats.edu.pk}}},
				\textbf{Shoaib Munir}$^{c}$  \footnote{E-mail: \texttt{\href{mailto:https:smunir@eaifr.org}{smunir@eaifr.org}}},
		 and \textbf{Umer Zubair}$^{d, e}$ \footnote{E-mail: \texttt{\href{mailto:umer@udel.edu}{umer@udel.edu}}} 	
		} 
		\vspace{.5cm}
		
		{\baselineskip 20pt \it
			$^{a}$ \it
			School of Mathematics and Physics, Hubei Polytechnic University, \\
			Huangshi 435003,
			China \\
			\vspace*{6pt}
				$^{b}$Department of Physics, COMSATS University Islamabad, Islamabad 45550, Pakistan \\
			\vspace*{6pt}
			$^{c}$ East African Institute for Fundamental Research (ICTP-EAIFR),University of Rwanda, Kigali, Rwanda \\
			\vspace*{6pt}
			$^d$Department of Physics, Saint Joseph's University, Philadelphia, PA 19131, USA\\
			\vspace*{6pt}
			$^e$Division of Science and Engineering, \\Pennsylvania State University, Abington, PA 19001, USA\\
						\vspace{2mm} }

		\vspace{1cm}
	\end{center}


\begin{abstract}
We analyse the shifted hybrid inflation in a no-scale supersymmetric $SU(5)$ GUT model which naturally circumvents the monopole problem. The no-scale framework is derivable as the effective field theory of the supersymmetric (SUSY) compactifications of string theory, and yields a flat potential with no anti-de Sitter vacua, resolving the $\eta$ problem. The model predicts a scalar spectral tilt $n_s$ compatible with the most recent measurements by the Planck satellite, while also accommodating large values of the tensor-to-scalar ratio $r$ ($\sim 0.0015$), potentially measurable by the near-future experiments. Moreover, the proton decay lifetime in the presence of the dimension-5 operators is found to lie above the current limit imposed by the Super-Kamiokande experiment. A realistic scenario of reheating and non-thermal leptogenesis is employed, wherein the reheating temperature $T_r$ lies in the $(2 \times 10^6 \lesssim T_r \lesssim 2 \times 10^9)$ GeV range, and at the same time realizing gravitino as a viable dark matter (DM) candidate.

\end{abstract}
\section{\large{\bf Introduction}}%

Inflation provides a successful phenomenological framework for addressing cosmological puzzles such as the size, the age, the homogeneity and the (approximate) geometrical flatness of the Universe \cite{reviews}. It also explains large-scale structures, the smallness of the primordial density perturbations measured in the cosmic microwave background (CMB), as well as the small tilt ($n_{s}=0.965$) in it and the inconsistency of this CMB with the Gaussian white-noise spectrum \cite{planck18}. The success of inflation has motivated several attempts to relate it to the Standard Model (SM) of particle physics, and at the same time to a candidate quantum theory of everything, including gravity, such as string theory. The characteristic energy scale of inflation is presumably intermediate between SM and quantum gravity, and inflationary models may thus provide a welcome bridge between the two of them.

A connection between inflation and a viable quantum theory of gravity at some very high scale and the SM at the electroweak (EW) scale makes models based on no-scale supergravity (SUGRA) a particularly attractive choice \cite{no-scale,Ellis:1983sf,LN}. No-scale SUGRA appears generically in models with ultraviolet completion using string theory compactifications \cite{Witten}. It is known to be the general form of the 4-dimensional effective field theory derivable from string theory that embodies low-energy supersymmetry. Moreover, no-scale SUGRA is an attractive framework for constructing models of inflation \cite{ENO8,EGNNO45} because it naturally yields a flat potential with no anti-de Sitter `holes', resolving the so-called $\eta$ problem. No-scale inflation also comfortably accommodates values of the $r$ and $n_s$ perfectly compatible with the most recent measurements of the Planck satellite, and potentially very similar to the values predicted by the original Starobinsky model \cite{Starobinsky:1980te}. For more detailed studies in the context of Starobinsky model, see Refs. \cite{ENO,CFKN,EKN,Einhorn:2009bh,Ferrara:2010in,Ferrara:2010yw,Ellis:2020lnc}. The simplest version of the Starobinsky proposal is equivalent to an inflationary model where the scalar field couples non-minimally to gravity. A natural choice for the inflaton field in the Grand Unified Theories (GUTs) with SUGRA is the Higgs field responsible for breaking the GUT gauge group. (for more details, see refs. \cite{Ahmed:2018jlv,Ahmed:2021dvo}).
  
Hybrid inflation is one of the most promising models of inflation that can be naturally realized within the context of SUGRA theories \cite{Dvali:1994ms,Copeland:1994vg,Linde:1997sj,Senoguz:2004vu,Rehman:2009nq,Buchmuller:2014epa,Abid:2021jvn}. In SUSY hybrid inflation, the scalar potential along the inflationary track is completely flat at the tree level. The inclusion of radiative corrections to this potential provides the necessary slope required to drive the inflaton towards the SM vacuum. In such a scenario \cite{Dvali:1994ms}, the CMB temperature anisotropy $\delta T/T$ is of the order $(M/m_{P})$, where $M$ is the breaking scale of the parent gauge group $G$ of the Universe at the start of inflation, and $m_P = 2.4 \times 10^{18}$ GeV is the reduced Planck mass. In order for the self-consistency of the inflationary scenario to be preserved, $M$ turns out to be comparable to the scale of grand unification, $M_{\text{GUT}}\sim10^{16}$ GeV, hinting that $G$ may be the GUT gauge group.
 
In the standard hybrid model of inflation, $G$ breaks spontaneously to its subgroup $H$ at the end of the inflation \cite{Dvali:1994ms,Rehman:2018nsn}, which leads to topological defects such as copious production of magnetic monopoles by the Kibble mechanism \cite{Kibble:1976sj}. These magnetic monopoles dominate the energy budget of the Universe, contradicting the cosmological observations. In the shifted \cite{Khalil:2010cp} or smooth \cite{Rehman:2014rpa,Rehman:2012gd,Ahmed:2022vlc} variants of the hybrid inflation, $G$ is instead broken during inflation, and in this way the disastrous monopoles are inflated away. 

In this article, we study shifted hybrid inflation formulated in the framework of no-scale SUGRA. In our model, the $SU(5)$ gauge group is spontaneously broken down to the $G_{SM}$ by the vacuum expectation value (VEV) of the $\Phi_{24}$ adjoint Higgs superfield. By generating a suitable shifted inflationary track wherein the $SU(5)$ is broken during inflation, the monopole density can be significantly diluted. The predictions of our model are consistent with the Planck's latest bounds \cite{planck18} on $n_s$ and $r$. Moreover, a wide range of the $T_r$ is obtained which naturally avoids the gravitino problem. A model of non-thermal leptogenesis via right-handed neutrinos is studied in order to explain the observed baryon asymmetry of the Universe (BAU). As compared to the $SU(5)$ model with shifted hybrid inflation studied outside the no-scale framework in \cite{Khalil:2010cp}, relatively large values of $r$ ($\sim 0.0015$) are obtained here, potentially measurable by the future experiments.

The layout of the paper is as follows. Sec. \ref{sec2} presents the basic description of the model including the field content, the superpotential and the global minima of the potential. The inflationary trajectories and the dimension-5 proton decay are discussed in Sec. \ref{sec3}. The inflationary setup and the theoretical details of reheating with non-thermal leptogensis are provided in Sec. \ref{sec4} and Sec. \ref{sec5}, respectively. The numerical analysis of the prospects of observing primordial gravity waves, and of leptogenesis and gravitino cosmology, is presented in Sec. \ref{sec6}. Finally, Sec. \ref{sec7} summarizes our findings. 

\section{\large{\bf The Supersymmetric $SU(5)$ Model with $U(1)_R$ Symmetry}} \label{sec2}

In the model we consider, the matter fields of the minimal supersymmetric standard model (MSSM) reside in the following representations of the $SU(5)$ supermultiplets.
\begin{equation}
\begin{split}
F_i &\equiv 10_i= Q_i\left(3,2, 1/6\right) + u_i^c\left(\overline{3},1, -2/3\right) + e^c \left(1,1,1\right), \\
\overline{f}_i &\equiv \overline{5}_i = d_i^c \left(\overline{3},1, 1/3 \right) + \ell_i \left(1,2, -1/2\right),     \\
\nu_i &\equiv 1_i =  \nu^c_i \left(1,1, 0\right)\,,
\end{split}
\end{equation}
where $i$ is the generation index $(i = 1,2,3)$ and $\nu^c_i$ are right-handed neutrnio superfields. The Higgs sector constitutes of a pair of 5-plet superfields \{$H \equiv H_5,\,\overline{H} \equiv \overline{H}_5$\} that contain the colour Higgs triplets and the two doublets of the MSSM, a gauge-singlet superfield $S$, and a 24-plet superfield $\Phi$ that belongs to the adjoint representation. This superfield $\Phi$ is responsible for breaking the $SU(5)$ gauge symmetry down to the SM gauge group by acquiring a non-zero VEV in the hypercharge direction. These Higgs superfields are decomposed under the $G_{SM}$ as
  \begin{eqnarray}\label{eq:Higgs-mul}
  H  &=& {H_T(3,1,-1/3)} + H_u (1,2,1/2),  \nonumber \\
\overline{H}  &=&  \overline{H}_T(\overline{3},1,1/3) + H_d(1,2,-1/2), \nonumber \\
 \Phi &=& \Phi_{24}(1,1,0) + W_H(1,3,0) + G_H(8,1,0)\nonumber \\
 &+& X_H(3,2,-5/6) + \overline{X}_H(\overline{3},2,5/6)\,.
 \end{eqnarray}
Following \cite{Khalil:2010cp}, the $U(1)_R$-charge assignments of the various superfields are 
\begin{equation}\label{eq:3}
R \left(S, \Phi, H, \overline{H}, F_{i}, \overline{f}_{i}, \nu_{i} \right) = \left(1, 0, \frac{2}{5}, \frac{3}{5}, \frac{3}{10}, \frac{1}{10}, \frac{1}{2} \right)\,.
\end{equation}

The $U(1)_R$-symmetric superpotential of the $SU(5)$ group containing the above superfields is written as
\begin{eqnarray} \label{eq:supotential}
W&=&\kappa S \left(\mu^2 - Tr(\Phi^2) - \beta \frac{Tr(\Phi^3)}{\kappa \, M_*}\right) +   \widetilde{\gamma} \overline{H}\Phi H + \widetilde{\delta} \overline{H} H  \nonumber \\
&+&y_{ij}^{(u)}F_i F_j H + y_{ij}^{(d,e)} F_i \overline{f}_j \overline{H} + y_{ij}^{(\nu)} \nu_i\overline f_{j} H + m_{ij}\nu_i \nu_j +\lambda_{ij}\frac{Tr(\Phi^2) \nu_i \nu_j}{M_*}\,,
\end{eqnarray}
where $\kappa$, $\widetilde{\gamma}$, and $\lambda_{ij}$ are dimensionless couplings, while $\widetilde{\delta}$ is a parameter with unit mass-dimension and $M_*$ is a high cut-off scale (compactification scale in a string model or the Planck scale) $M_{\text{GUT}} \lesssim M_* \lesssim m_{P}$. The Yukawa couplings $y_{ij}^{(u)}$, $y_{ij}^{(d,e)}$ and $y_{ij}^{(\nu)}$ in the second line above generate the quark and lepton masses after the EW symmetry-breaking, whereas the last two terms generate the right-handed neutrino masses, and are thus relevant for leptogenesis in the post inflationary era. 

\subsection{\bf The Global SUSY Minima} \label{sec2_1}

The first line of the superpotential in Eq. (\ref{eq:supotential}) contains the terms responsible for the shifted hybrid inflation, and can be rewritten in component form as
 \be%
 W \supset \kappa S\left( \mu^2 -\frac{1}{2}
 \sum_{i}\phi_{i}^{2} - \frac{\beta}{4\kappa \, M_*}d_{ijk}\phi_{i}\phi_{j}
 \phi_{k}\right)+\widetilde{\gamma} \, T_{a b}^{i}\phi^{i}\overline{H_{a}}H_{b} + \widetilde{\delta} \,
 \overline{H_a}H_{a}\,,%
 \label{eq:superpot-shift}%
 \ee %
where we have expressed the superfield $\Phi$ in the $SU(5)$ adjoint basis $\Phi= \phi_i T^i$ with $\tr(T_i T_j)=\delta_{ij}/2$ and $d_{ijk}=2 \tr(T_i\{T_j,T_k\})$. Here the indices $i$, $j$ and $k$ run from 1 to 24, whereas $a$ and $b$ run from 1 to 5. The global $F$-term potential is obtained from $W$ as
  \begin{eqnarray}
  V_{F}\supset \sum_{i} \left| \frac{\partial W}{\partial z_{i}} \right|^{2} &=&\sum_{i}\left|\kappa S \phi_{i}+\frac{3
 	\beta}{4 M_*}d_{ijk} S \phi_{j} \phi_{k}
 - \widetilde{\gamma} T_{a
 	b}^{i}\overline{H_{a}}H_{b}\right|^{2}
+ \sum_{b}\left|\widetilde{\gamma}
 T_{a b}^{i}\phi^{i}\overline{H_{a}}+\widetilde{\delta}
 \overline{H_{b}}\right|^{2} \nonumber \\
&+&  \kappa^2 \left| \mu^2-\frac{1}{2}
 \sum_{i}\phi_{i}^{2}-\frac{\beta}{4
 	M_*}d_{ijk}\phi_{i}\phi_{j}
 \phi_{k} \right|^{2}
+\sum_{b}\left|\widetilde{\gamma} T_{a
 	b}^{i}\phi^{i}H_{a}+\widetilde{\delta}
 H_{b}\right|^{2}\,,%
 \label{eq:scalarpot-shift}
 \end{eqnarray}%
where $z_i$ are the scalar components of the Higgs superfields. The global SUSY minimum of the above potential lies at the following VEVs of the fields
  \begin{eqnarray} \label{global_minima}
  \left< S \right> = \left<H_a\right> = \left<\overline{H_a}\right> = 0, 
  \end{eqnarray}
  while $\langle \phi_{i} \rangle$ satisfy the condition
  \begin{equation}
  \sum_{i=1}^{24}\left<\phi_{i}\right>^{2} + \frac{\beta}{2 \kappa \, M_*}
  d_{ijk}
  \left<\phi_{i}\right> \left<\phi_{j}\right> \left<\phi_{k}\right> =2 \mu^{2}\,. 
  \label{phi_vev}
  \end{equation}
  The VEV matrix $\left<\Phi_{i}\right> = \left<\phi_{i}\right> T^i$ can be aligned in the hypercharge ($i = 24$) direction using the $SU(5)$ transformation
  \begin{equation}
  \left< \Phi_{24} \right> = \frac{\left< \phi_{24} \right>}{\sqrt{15}} \left( 1, 1, 1, - 3/2, - 3/2 \right)\,,
  \end{equation} 
  such that $\left< \phi_i \right> = 0, \, \forall \, i \neq 24$ and $\left< \phi_{24} \right> \equiv \upsilon/\sqrt{2} $, where $d_{24\,24\,24} = -1/\sqrt{15}$ and $\upsilon$ satisfies
  \begin{equation}
  4 \mu^2 =  \upsilon ^2 - \frac{\beta}{2 \sqrt{30} \kappa \, M_*} \upsilon ^3\,. \label{full_vev}
  \end{equation}
 The $D$-term contribution to the potential,
  \begin{equation}
  	V_D = \frac{g_{5}^2}{2} \sum_{i} \left( f^{ijk} \phi_j \phi_{k}^{\dagger}  + T^i \left( \left| H_a \right|^2  - \left| \overline{H}_a \right|^2  \right) \right)^2\,,
  \end{equation}
  also vanishes for $\phi=\phi^{\ast}$ and $\vert \overline{H}_a \vert =\vert H_a \vert$\,.
  
\section{\large{\bf Inflationary Trajectories}} \label{sec3}
\begin{figure}[t]
	\centering 
	\begin{subfigure}{0.5\textwidth}
		\centering 
		\caption{Standard ($\alpha = 0$)}
		\includegraphics[width=1\linewidth]{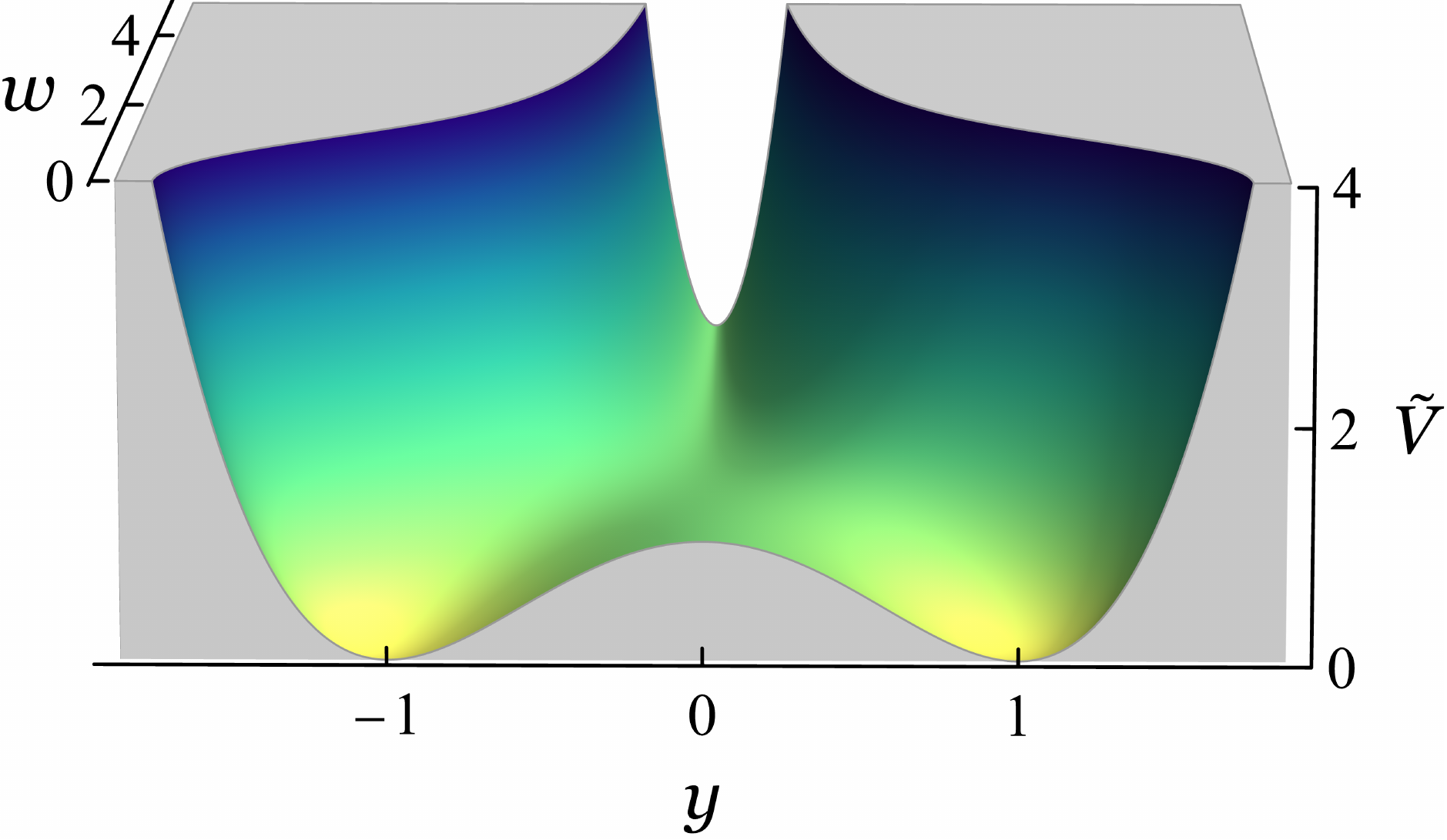}
		\label{panel_a}
	\end{subfigure}
	
	\begin{subfigure}{0.496\textwidth}
		\caption{Shifted ($\alpha = 0.25$)}
		\centering \includegraphics[width=1\linewidth]{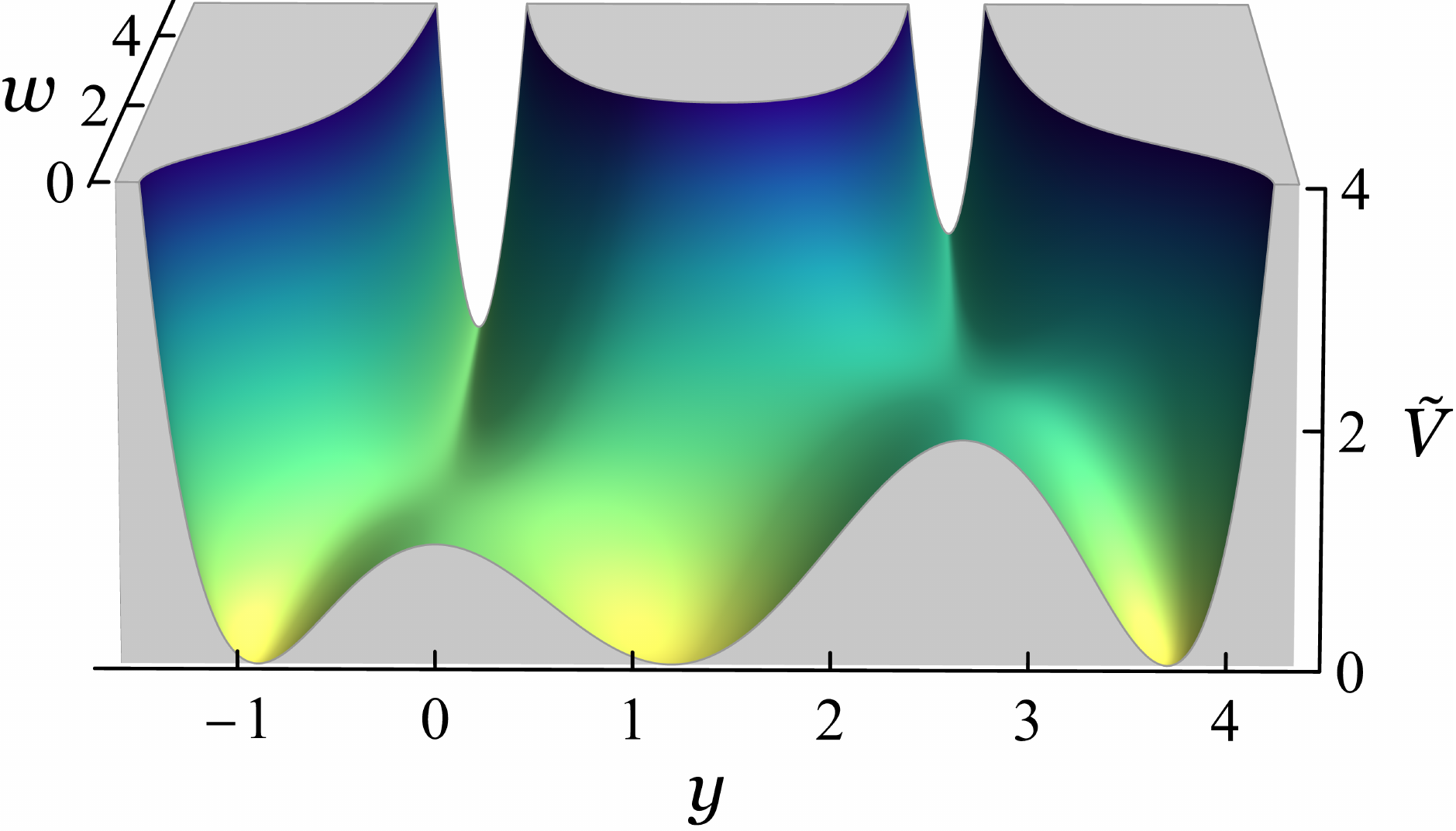}
		\label{panel_b}
	\end{subfigure}
	\begin{subfigure}{0.496\textwidth}
		\caption{Shifted ($\alpha = 0.3$)}
		\centering \includegraphics[width=1\linewidth]{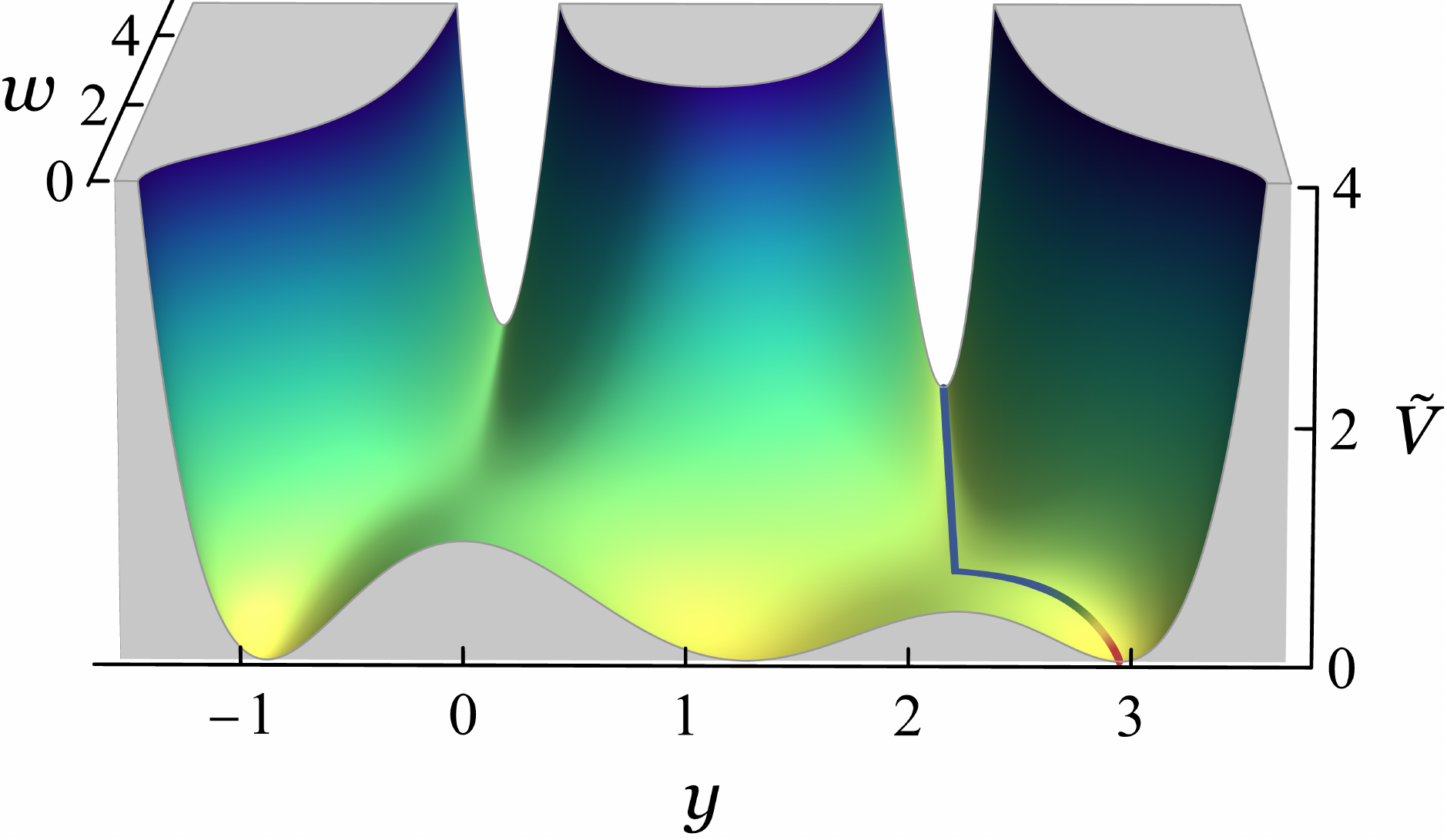}
		\label{panel_c}
	\end{subfigure}
	\caption{Normalized scalar potential $\widetilde{V}(w, y) = V (w, y) / \kappa^2\mu^4$ as a function of dimensionless variables $w$ and $y$, for different values of parameter $\alpha$. The standard hybrid inflation potential is reproduced for $\alpha = 0$ in panel (a). For $\alpha \neq 0$, an additional shifted trajectory appears which lies higher than the standard trajectory for $\alpha < \sqrt{2/27}$ and lower than the standard trajectory for $\alpha > \sqrt{2/27}$. These shifted hybrid inflation potentials are shown in panels (b) and (c) for $\alpha = 0.25$ and $\alpha = 0.3$, respectively.}
	\label{fig1}
\end{figure}

The scalar potential in Eq.~(\ref{eq:scalarpot-shift}) can be rewritten in terms of the dimensionless variables%
  \be %
  y = \frac{\phi_{24}/\mu}{\sqrt{2}}~, ~~~~~~~~~~~~
  w = \frac{S/\mu}{\sqrt{2}}\,, %
  \ee %
as
\be
\widetilde{V}(w, y)=\frac{V (w, y)}{\kappa^2\mu^4}=\left(1-y^2+\alpha y^3\right)^2+2 w^2 y^2\left(1-\frac{3\alpha y}{2}\right)^2\, , \label{dimless_potential}
\ee
with  $\alpha = \beta \mu/ \sqrt{30}\, \kappa M_*$. This dimensionless potential exhibits the following three extrema
\begin{equation}
y_1=0, \label{1st minima}
\end{equation}
\begin{equation}
y_2=\frac{2}{3 \alpha }, \label{2nd minima}
\end{equation}
\begin{eqnarray}
\text{and}\quad y_3 = \frac{1}{3\alpha} &+& \frac{1}{3 \sqrt[3]{2} \alpha }\Bigg( \sqrt[3]{2 -27 \alpha ^2+\sqrt{\left(2-27 \alpha ^2\right)^2+4 \left(9 \alpha ^2 w^2-1\right)^3}}   \nonumber\\ 
&-&    \sqrt[3]{-2 + 27 \alpha ^2+\sqrt{\left(2-27 \alpha ^2\right)^2+4 \left(9 \alpha ^2 w^2-1\right)^3}} \Bigg)\,,
\label{3rd minima}
\end{eqnarray}
for any constant value of $w$. The dimensionless potential $\widetilde{V}(w, y)$ is displayed in Fig. \ref{fig1} for different values of $\alpha$. The first extremum $y_1$ with $\alpha = 0$ corresponds to the standard hybrid inflation for which $\{y = 0$,\,$w > 1\}$ is the only inflationary trajectory that evolves at $w = 0$ into the global SUSY minimum at $y = \pm 1$ (panel (a)). For $\alpha \neq 0$, a shifted trajectory appears at $y = y_2$, in addition to the standard trajectory at $y = y_1 =0$, which is a local maximum (minimum) for $w < \sqrt{4/27\alpha^2-1}$ ($w > \sqrt{4/27\alpha^2-1}$). For $\alpha < \sqrt{2/27}\simeq 0.27$, this shifted trajectory lies higher than the standard trajectory (panel (b)). In order to have suitable initial conditions for realizing inflation along the shifted track, we assume $\alpha > \sqrt{2/27}$, for which the shifted trajectory lies lower than the standard trajectory (panel (c)). Moreover, to ensure that the shifted inflationary trajectory at $y_2$ can be realized before $w$ reaches zero, we require $\alpha < \sqrt{4/27} \simeq 0.38$. Thus, for $0.27 < \alpha < 0.38$, while the inflationary dynamics along the shifted track remain the same as for the standard track, the $SU(5)$ gauge symmetry is broken during inflation, hence alleviating the magnetic monopole problem. As the inflaton slowly rolls down the inflationary valley and enters the waterfall regime at $w =\sqrt{4/27\alpha^2-1}$, its  fast rolling ends the inflation, and the system starts oscillating about the vacuum at $w= 0$ and $y=y_3$\,.

\subsection{\bf Dimension-5 Proton Decay} \label{sec2_3}
\begin{figure}[t]
	\centering
	\includegraphics[width=9cm]{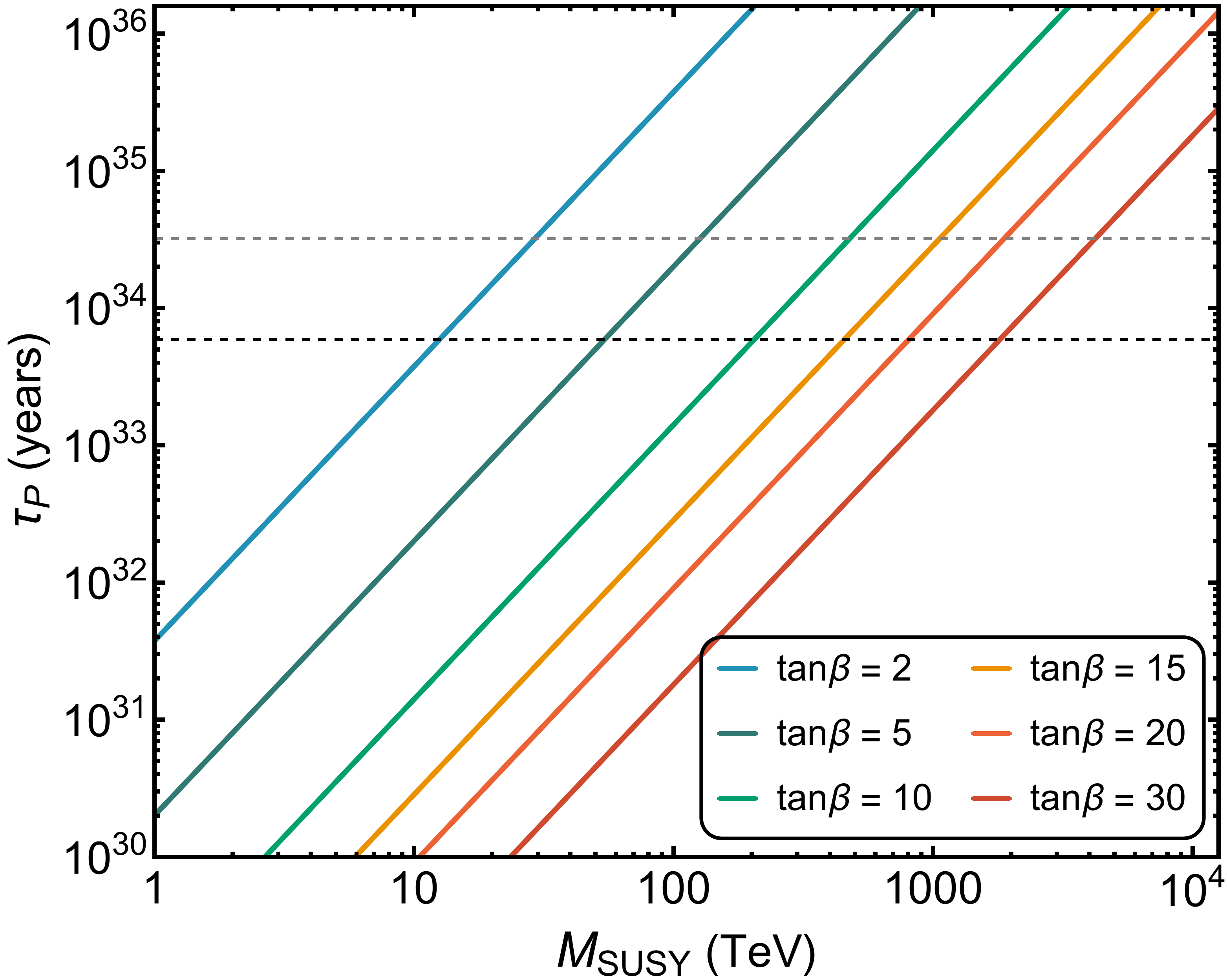}
	\caption{The proton lifetime for the decay $p \rightarrow K^+ \bar{\nu}$ as a function of SUSY breaking scale $M_{\text{SUSY}}$ for different values of $\tan \beta$. The curves are drawn for $SU(5)$ symmetry breaking scale $M_{\alpha}$ fixed at $2 \times 10^{16}$ GeV with $\alpha = 0.3$. The black and gray dashed lines represent the Super-Kamiokande ($\tau_p = 5.9 \times 10^{33}$ years) bounds and future Hyper-Kamiokande expected bounds ($\tau_p = 3.2 \times 10^{34}$ years)  on proton lifetime, respectively.}
	\label{fig:proton_decay}
\end{figure}
Upon breaking of the $SU(5)$ symmetry, the last two terms in the $W$ in Eq. (\ref{eq:superpot-shift}) can be rewritten as,
\begin{equation} \label{doublet_triplet_splitting}
	W_H  \supset \left(\widetilde{\delta} - \frac{3 \widetilde{\gamma} \phi_{24}^0}{2 \sqrt{30}}\right) {H}_{u} H_d +  \left(\widetilde{\delta} + \frac{ \widetilde{\gamma} \phi_{24}^0}{\sqrt{30}}\right) \overline{H}_{T}  H_T  
	\equiv \mu_H {H}_{u} H_d +  M_{H_T} \overline{H}_{T} H_T\,,
\end{equation}
where $\mu_H$ is identified with the usual MSSM $\mu$-parameter, which is taken to be of the order of TeV scale, and $M_{H_T}$ is the color-triplet mass parameter given by
\begin{eqnarray}
	M_{H_T} \simeq \frac{10 \widetilde{\gamma} M_{\alpha}}{3 \sqrt{30\left(\frac{4}{27} - \alpha^2\right)}}\,.
	\label{eq:color_triplet_higgs_mass}
\end{eqnarray}
The above mass-splitting between the Higgs doublet and triplet can be addressed by fine tuning of the parameters $\tilde{\delta}$ and $\tilde{\gamma}$, such that
\begin{equation*}
	\widetilde{\delta} \simeq \frac{3 \widetilde{\gamma} \phi_{24}^0}{2 \sqrt{30}}\,.
\end{equation*}
However, the fermionic components of $H_T$, the color-triplet Higgsinos, contribute to the proton decay via a dimension-5 operator, which typically dominates the gauge boson mediated dimension-6 operators. 

The proton lifetime for the decay $p \rightarrow K^+ \bar{\nu}$ can be approximated by the following formula \cite{Nagata:2013ive},
\begin{equation}
	\tau_p \simeq 4 \times 10^{31} \times \sin^4 2\beta \left( \frac{M_{\text{SUSY}}}{1~ \text{TeV}} \right)^2 \left( \frac{M_{H_T}}{10^{16} ~ \text{GeV}} \right)^2 \text{yrs}\,,
	\label{proton_lifetime}
\end{equation}
The proton lifetime $\tau_p$ is shown in Fig. \ref{fig:proton_decay} as a function of $M_{\text{SUSY}}$ for different values of $\tan \beta$, using Eq. \eqref{full_vev}, \eqref{eq:color_triplet_higgs_mass} and \eqref{proton_lifetime}.  The curves are drawn for the $SU(5)$-breaking scale $M_{\alpha}$ fixed at $2 \times 10^{16}$ GeV, with $\alpha = 0.3$. It can be seen that the proton lifetime is consistent with the experimental bound, $\tau_p > 5.9 \times 10^{33}$ years, from Super-Kamiokande \cite{Super-Kamiokande:2016exg}, for $M_{\text{SUSY}} \gtrsim 12.5$ TeV and can be observed by Hyper-Kamiokande \cite{Hyper-Kamiokande:2018ofw}. Our model thus remains safe from proton decay even while adequately addressing the doublet-triplet splitting problem.

\section{\large{\bf No-Scale Shifted Hybrid Inflation}} \label{sec4}

The K\"ahler potential with a no-scale structure, after including contributions from the relevant fields in the model, takes the following form
\begin{equation}
K=-3 m_P^2 \log \Delta\,,
\label{KahlerPotential}
\end{equation} 
with
\be
\Delta=T + T^{\ast}-\frac{\tr(\Phi^2)+S^\dagger S+\overline{H}  H+ \nu^{c} \nu^{c}}{3 m_P^2}+\gamma\frac{(S^\dagger S)^2}{3 m_P^4}+\zeta\frac{\tr(\Phi^4)}{3 m_P^4}+\sigma\frac{(S^\dagger S)^3}{3 m_P^6}\,,
\ee 
where $\gamma$, $\zeta$ and $\sigma$ are dimensionless couplings, and $T$ and $T^{*}$ are K\"ahler complex moduli fields given as $T=\left(u+i\, v\right)$, so that $T+ T^{\ast}=2u$ with $u=1/2$. The $F$-term SUGRA scalar potential is given by
\begin{equation}
V_{F}=e^{K/m_{P}^2} \left[\left(K_{i\bar{j}} \right)^{-1}\left(D_{z_{i}}W\right)\left(D_{z_{j}}W\right)^{*}-\frac{3\arrowvert W\arrowvert^{2}}{m_{P}^{2}}\right]\,, \label{Einstein frame SUGRA potential}
\end{equation}
where we have defined
\begin{equation}
D_{z_{i}}W\equiv \frac{\partial W}{\partial z_{i}}+\frac{\partial K}{\partial z_{i}}\frac{W}{m_{P}^{2}}, \ \ \ \ K_{i\bar{j}}\equiv \frac{\partial^{2}K}{\partial z_{i}\partial z_{j}^{*}}\,,
\end{equation}
and $D_{z_{i}^{*}}W^{*}=\left( D_{z_{i}}W\right)^{*}.$

Since SUSY is temporarily broken along the inflationary trajectory, the radiative corrections to the above $V_F$ along with the soft SUSY-breaking potential $V_{\text{Soft}}$ can lift its flatness, while also providing the necessary slope for driving inflation. For a detailed discussion on the mass spectrum of the model, see Ref. \cite{Khalil:2010cp}. The effective contribution of the one-loop radiative corrections can be calculated using the Coleman-Weinberg formula \cite{Coleman-Weinberg: 1973} as
\be
V_{1\text{-loop}}=\kappa^2 M_{\alpha}^{4}\left(\frac{\kappa^2}{16\pi^2}[F(M_\alpha^2,x^2)+11\times 25 F(5M_\alpha^2,5x^2)]\right)\,, \label{loop corrections}
\ee
where, for a non-canonically normalized field $x\equiv |S|/M_\alpha$,
\be
F(M_\alpha^2,x^2)=\frac{1}{4}\left((x^4+1)~\text{ln}\left(\frac{x^4-1}{x^4}\right)+2x^2~\text{ln}\frac{(x^2+1)}{(x^2-1)}+2~\text{ln}\left(\frac{\kappa^2 M_\alpha^2 x^2}{Q^2}\right)-3\right)\,,
\ee
with $M_\alpha^2=\mu^2\left(\frac{4}{27\alpha^2}-1\right)$, and $Q$ being the renormalization scale. 

As for the breaking of SUSY, in this study we consider the scenario wherein it is communicated gravitationally from the hidden sector to the observable sector. Following \cite{Nilles:1983ge}, the soft SUSY-breaking potential thus reads
\begin{equation}
V_{\text{Soft}}\simeq a m_{3/2} \kappa M_{\alpha }^3 x + M_S^2 M_{\alpha }^2 x^2 +\frac{8 M_\phi^2 M_{\alpha }^2}{9 \alpha^2\left(4/27\alpha^2 - 1\right)}\,,  \label{SHIpot} 
\end{equation}
with
\begin{equation*}
	a=2\vert A-2 \vert \cos \left(\arg S + \arg \vert A-2 \vert\right)\,.
\end{equation*}
Here 
$m_{3/2}$ is the gravitino mass, $A$ is the complex coefficient of the trilinear soft SUSY-breaking terms, $a$ and $M_S$ are the coefficients of the soft linear and mass terms for $S$, respectively, while $M_\phi$ is the soft mass parameter for the $\phi$ field. The complete effective scalar potential during inflation is then given as
\begin{eqnarray}
V(x) &\simeq& V_{F}+V_{1\text{-loop}}+V_{\text{Soft}} \nonumber \\
&\simeq& \kappa ^{2}M_{\alpha}^{4}\left[\left( 1 + 2 \left( \left(2 \gamma - \frac{1}{3}\right) x^2 - \frac{1}{\left(1 - 27 \alpha^2 / 4\right)} \right)\left(\frac{M_{\alpha}}{m_P}\right)^2   \right.\right.\nonumber \\
&+& \left.\left. \left(\frac{16\left(1-2\zeta\right)}{3\left(4-27\alpha^2\right)^2}+\frac{8x^2\left(1-12\gamma\right)}{3\left(4-27\alpha^2\right)} -x^4\left(16\gamma^2+\frac{14}{3}\gamma+\frac{9\sigma}{2}-\frac{5}{9}\right) \right) \left(\frac{M_{\alpha}}{m_P}\right)^4 \right) \right.\nonumber \\
&+& \left. \frac{\kappa^2}{16 \pi^2} \left( F(M_{\alpha}^2,x^2)
+11\times 25\,F(5M_{\alpha}^2,5\,x^2) \right)\right]  \nonumber \\
&+&  a m_{3/2} \kappa M_{\alpha }^3 x + M_S^2 M_{\alpha }^2 x^2 +\frac{8 M_\phi^2 M_{\alpha }^2}{9 \alpha^2\left(4/27\alpha^2 - 1\right)}\,,
  	\label{dterm}
  \end{eqnarray}
where the $F$-term scalar potential along the shifted trajectory in the $D$-flat direction has been obtained from Eq. (\ref{Einstein frame SUGRA potential}). Finally, the action of our model is given by
 \begin{equation}
\begin{split}
\mathcal{A}= \int dx^{4}\sqrt{-g}\left[\frac{m^2_{p}}{2}\mathcal{R}- K^{i}_{j}\partial_{\mu} x^{i}\partial^{\mu} x^{j} -V({x})\right]\,,
\end{split}
\end{equation}
where $\mathcal{R}$ is the Ricci scalar. Introducing a canonically normalized field $z$ satisfying
\begin{eqnarray}
\left(\frac{dz}{dx}\right)^2=\frac{\partial^2K}{\partial S^{\dagger} \partial S} \quad \text{with}\quad S^{\dagger}= S=x M_{\alpha} \quad \text{and}\quad \phi_{24}=\frac{\sqrt{2}M_{\alpha}y_{2}}{\left(4/27\alpha^2-1\right)}\,,
\end{eqnarray}
requires modification of the slow-roll parameters, as shown later in section \ref{sec6}.

\section{\large{\bf Reheating with Non-thermal Leptogenesis }}\label{sec5}

A complete inflationary scenario should be followed by a successful reheating that satisfies the constraint, $T_r<10^9$ GeV, from gravitino cosmology, and generates the observed BAU. At the end of the inflationary epoch, the system falls towards the SUSY vacuum and undergoes damped oscillations about it. The inflaton (oscillating system) consists of two complex scalar fields $S$ and $\theta=(\delta \phi+\delta \overline{\phi})/\sqrt{2}$. The canonical normalized inflaton field can be defined as
\begin{equation}
\delta \widetilde{ \phi} =  \langle J_{0} \rangle \delta \phi\,,
\end{equation} 
with
\begin{equation}
	\delta \phi = \phi - \langle \phi_{24} \rangle _{y_{3}}, \quad \text{and} \quad J_{0}\equiv \frac{dz}{dx}\Big\vert_{\text{Minimum}}=\left(1-\frac{\langle \phi_{24} \rangle^2 _{y_{3}}}{6 m_P^2} + \frac{\zeta \langle \phi_{24} \rangle^4_{y_{3}}}{12 m_P^4}\right)^{-1/2}\,, 
\end{equation}
where
\begin{equation}
\langle \phi_{24} \rangle _{y_{3}}\equiv\frac{\sqrt{2}M_{\alpha}y_{3}}{\left(4/27\alpha^2-1\right)}\,.
\end{equation} 

The decay of the inflaton field, with its mass given by 
\begin{eqnarray}
\widetilde{m}_{\text{inf}}^2&=& \frac{d^2V(\phi)}{dz^2}= \dfrac{1}{J_{0}}\left(\frac{d^2V(\phi)}{d\phi^2}\right)= \dfrac{m_{\text{inf}}^2}{J_{0}}=\frac{2\kappa^2 M_{\alpha}^2y_{3}^2\left(1-3\alpha y_{3} /2 \right)^2}{J_{0}\left(4/27 \alpha^2 - 1\right)}\,,
\end{eqnarray}
gives rise to the radiation in the Universe. The inflaton decay into the higginos and the right-handed neutrinos is induced by the superpotential terms,
\begin{equation}
W \supset \widetilde{\gamma} \overline{H}\Phi H+\lambda_{ij}\frac{\tr(\Phi^2) \nu_i \nu_j}{M_*}\,.
\label{Infnu1}
\end{equation}
The Lagrangian terms relevant for inflaton decay into neutrinos are\cite{Endo:2007sz,Pallis:2011gr},
\begin{eqnarray}
\mathcal{L}_{\text{inf}\rightarrow \nu_{i}^{c} \nu_{i}^{c}} &=& 
- \frac{1}{2} e^{K/2m_{P}^2} 
\left(
K_\phi W_{\phi,\nu_{i}^c\nu_{j}^c} + W_{\phi ,\nu_{i}^c\nu_{j}^c}
- 2 \Gamma^{\nu_{i}^c}_{\phi ,\nu_{j}^c} W_{\phi,\nu_{i}^c\nu_{j}^c}
\right)\, \phi \nu_{i}^c\nu_{i}^c + {\rm h.c.} \nonumber\\
&\to& -g_{\nu_{ij}^c}  \left(\delta\widetilde{ \phi} \nu_{i}^{c} \nu_{j}^{c}+h.c\right)+...\,,
\label{eq:int-1}
\end{eqnarray}
where $W_{\phi,\nu^c_i \nu^c_j}$ is the second derivative of $W$ with respect to $\nu^c$, and $g_{\nu_{ij}^c}$ is the effective inflaton-neutrino coupling, defined as
\begin{equation}
g_{\nu_{ij}^c}=\frac{\sqrt{2}\lambda_{ij} M_{\alpha } y_{3}J_{0}^4}{M_*\left(4/27\alpha^2 - 1\right)^{1/2}}\,.
\end{equation} 

The decay width in the flavor-diagonal basis is thus given by \cite{Lazarides:1998qx}
\begin{eqnarray}
\Gamma_{\text{inf}\rightarrow \nu^{c} \nu^{c}}&=&\dfrac{1}{64\pi}g_{\nu_{i}^c}^2 \widetilde{m}_{\text{inf}}\left(1-\frac{4M_{\nu_{i}^c}^2}{\widetilde{m}_{\text{inf}}^2}\right)^{3/2}, \nonumber\\
&=&\dfrac{J_{0}^8}{16\pi}\widetilde{m}_{\text{inf}}\left(\dfrac{M_{\nu_{i}^{c}}}{ \quad \langle \phi_{24} \rangle _{y_{3}}}\right)^2\left(1-\frac{4M_{\nu_{i}^c}^2}{\widetilde{m}_{\text{inf}}^2}\right)^{3/2}\,,
\end{eqnarray}
where $M_{\nu_i^c}$ represent the eigenvalues of the soft neutrino mass matrix. Note that this term violates the lepton number by two units, $\Delta L = 2$. In addition, the Dirac mass terms for the neutrinos are obtained from $W\supset y_{ij}^{(\nu)}\,\nu_{i}^c\,\bar{f}_j\,H \to {m_{\nu_D}}_{ij}\nu_i\nu_j^c$ upon EW symmetry-breaking. The small neutrino masses, consistent with the results from the neutrino oscillation experiments, are obtained by integrating out the heavy right-handed neutrinos, so that
\begin{equation}
{m_{\nu_D}}_{\alpha\beta}=-\sum_{i}{y^{(\nu)}}_{i\alpha}{y^{(\nu)}}_{i\beta}\frac{v_{u}^2}{M_{\nu_i^c}}\,.
 \label{mneu1}
\end{equation}
The Dirac mass matrix above can be diagonalised by a unitary matrix $U_{\alpha i}$ as ${m_{\nu_D}}_{\alpha\beta} =  U_{\alpha i} U_{\beta i} m_{\nu_D}$, with $m_{\nu_D} = {\rm diag}(m_{\nu_{1}}, m_{\nu_{2}}, m_{\nu_{3}})$\,. 

The Lagrangian relevant for the inflaton decay to $H$ and $\overline{H}$ is
\begin{eqnarray}
\mathcal{L}_{\text{inf}\rightarrow H \overline{H}} &=& 
- \frac{1}{2} e^{K/2m_{P}^2} 
\left(
K_\phi W_{H \overline{H}} + W_{\phi H \overline{H}}
- 2 \Gamma^H_{\phi \overline{H}} W_{H \overline{H}}
\right)^*m_{\text{inf}}\, 
\phi H \overline{H} + {\rm h.c.} \nonumber\\
&\to& -g_{H} \widetilde{m}_{\text{inf}} \left(\delta \widetilde{\phi} H \overline{H}+ h.c\right)+...\,,
\label{eq:2scalar}    
\end{eqnarray} 
with the effective coupling 
\begin{equation}
	g_{H}=\dfrac{ \widetilde{\gamma} }{2}J_{0}^{3}\,.
\end{equation}
This leads to the partial decay width \cite{Lazarides:1998qx},
\begin{eqnarray}
\Gamma_{\text{inf}\rightarrow  H \overline{H}}=\frac{g_{H}^2}{8\pi}\widetilde{m}_{\text{inf}}\,.
\end{eqnarray}

The reheating temperature of the Universe depends on the combined decay width $\Gamma=\Gamma_{\text{inf}\rightarrow \nu_{i}^{c} \nu_{i}^{c}}+\Gamma_{\text{inf}\rightarrow  \overline{H} H}$ as
\begin{eqnarray}
T_{r}=\left(\frac{72}{5\pi^2 g^{\ast}}\right)^{1/4}\sqrt{\Gamma m_{P}}\,.
\end{eqnarray}
Assuming a standard thermal history, the number of $e$-folds, $N_{0}$, can be written in terms of the $T_r$ as \cite{Liddle:2003as}
\begin{equation}\label{n0}
N_0=54+\frac{1}{3}\ln\Big(\frac{T_r}{10^9\text{ GeV}}\Big)+\frac{2}{3}\ln\Big(\frac{V(x)^{1/4} }{10^{15}\text{ GeV}}\Big)\,.
\end{equation}
The ratio of the lepton number density to the entropy density in the limit $T_r < M_{\nu_1^c}\leq m_{\text{inf}} /2 \leq M_{\nu_{2,3}^c}$ is defined as
\begin{equation}
\frac{n_{L}}{s}\sim \frac{3}{2}\frac{\Gamma_{\text{inf}\rightarrow \nu_{1}^{c} \nu_{1}^{c}}}{\Gamma}\frac{T_{r}}{\widetilde{m}_{\text{inf}}}\epsilon_{cp}\,,
\end{equation}
where $\epsilon_{cp}$ is the CP-asymmetry factor, which is generated from the out-of-equilibrium decay of the lightest right-handed neutrino. For a normal hierarchical pattern of the light neutrino masses, this factor becomes \cite{Hamaguchi:2002vc} 
\begin{equation}
\epsilon_{cp} = \frac{3}{8\pi}\frac{M_{\nu_1^c} m_{\nu_{3}}}{v_{u}^2}\delta_{\rm eff}\,, 
\end{equation}
where $m_{\nu_3}$ is the mass of the heaviest light neutrino, $v_{u}$ is the VEV of the Higgs doublet $H_u$, and $\delta_{\rm eff}$ is the CP-violating phase. 

The lepton asymmetry from experimental observations is \cite{Planck:2018vyg}, 
\begin{eqnarray}
\mid n_L/s\mid\approx\left(2.67-3.02\right)\times 10^{-10}\,. 
\end{eqnarray}
In the numerical estimates discussed below, we take $m_{\nu_3} = 0.05$ eV, $|\delta_{\rm eff}|\le1$, $v_u = 174$ GeV, while assuming large $\tan \beta$. A successful baryogenesis is usually generated through the sphaleron process \cite{Kuzmin:1985mm,Fukugita:1986hr}, where an initial lepton asymmetry, given by 
\begin{equation}
\frac{n_L}{s} \lesssim 3 \times 10^{-10} \left(\frac{\Gamma_{\text{inf}\rightarrow \nu^{c} \nu^{c}}}{\Gamma}\right)\left(\frac{T_r}{\widetilde{m}_{\text{inf}}}\right)\left(\frac{M_{\nu_1^c}}{10^6 \text{ GeV}}\right)\left(\frac{m_{\nu_3}}{0.05 \text{ eV}}\right)\delta_{\rm eff},
\label{fin1}
\end{equation}
is partially converted into baryon asymmetry as $n_{B}/s=-0.35n_L/s$.
 
\section{Numerical analysis} \label{sec6}
 
\subsection{Inflationary Predictions} \label{sec6_1}
The inflationary slow-roll parameters can be expressed as
\begin{gather}
	\epsilon=\dfrac{1}{4}\left(\frac{m_{P}}{M_{\alpha}}\right)^2\left(\frac{V^{\prime}\left(x\right)}{V(x)z^{\prime}\left(x\right)}\right)^{2},,\quad \eta=\dfrac{1}{2}\left(\frac{m_{P}}{M_{\alpha}}\right)^2\left(\frac{V^{\prime\prime}\left(x\right)}{V(x)\left(z'\left(x\right)\right)^{2}}-\frac{V^{\prime}\left(x\right)z''\left(x\right)}{V(x)\left(z^{\prime}\left(x\right)\right)^{3}}\right)\,, \nonumber\\
	\begin{aligned}
		\text{and}\quad s^{2}=&~\dfrac{1}{4}\left(\frac{m_{P}}{M_{\alpha}}\right)^4\left(\frac{V^{\prime}\left(x\right)}{V(x)z'\left(x\right)}\right)\Bigg(\frac{V^{\prime\prime\prime}\left(x\right)}{V(x)\left(z'\left(x\right)\right)^{3}} - 3\frac{V^{\prime\prime}\left(x\right)z'^{\prime}\left(x\right)}{V(x)\left(z'\left(x\right)\right)^{4}} \\
		+&~3\frac{V^{\prime}\left(x\right)\left(z''\left(x\right)\right)^{2}}{V(x)\left(z'\left(x\right)\right)^{5}}-\frac{V^{\prime}\left(x\right)z'''\left(x\right)}{V(x)\left(z'\left(x\right)\right)^{4}}\Bigg)\,,
	\end{aligned}
	\label{eq:infpar}
\end{gather}
\noindent where a prime denotes a derivative with respect to $x$. In terms of these parameters, we obtain
\begin{equation}
	r\simeq16 \epsilon,\quad n_{s}\simeq 1+2\eta-6\epsilon, \quad \text{and}\quad \frac{dn_{s}}{d\ln k}\simeq 16\epsilon\eta-24\epsilon^{2}+2s^{2}\,,
	\label{eq:r-and-ns}
\end{equation}
where the last quantity gives the running of $n_s$. The number of $e$-folds is given by
\begin{equation}
	N_{0}=2\left(\frac{M_{\alpha}}{m_{P}}\right)^2\int_{x_{e}}^{x_{0}}\left(\frac{V\left(x\right)z'(x)^{2}}{V^{\prime}(x)}\right) dx\,,
\end{equation}
where $x_{0}$ is the field value at the pivot scale and $x_e$ is the field value at the end of inflation (i.e., when  $\epsilon = 1$). Finally, the amplitude of curvature perturbation $\Delta_{R}$ is obtained as
\begin{equation}
	\Delta_{R}^{2}=\frac{V\left(x\right)}{24 \pi^{2} \epsilon\left(x\right)}\,.
\end{equation}

The results of our numerical calculations are displayed in Figs. \ref{fig:results1} and \ref{fig:results2}, which show the ranges of $\kappa$, $S_0$, and $T_r$ in the $\tilde{\gamma}-\sigma$ plane. The color map in Fig. \ref{fig:results1} corresponds to $r$ in the top two panels and to the coupling $\gamma$ in the bottom panel. The light-shaded region in all the panels implies that the inflaton predominantly decays into the neutrinos, whereas the dark-shaded region represents a Higgsino-dominant decay. In obtaining these results, we have used up to second-order approximation on the slow-roll parameters, and have set $\zeta = 0$, $\alpha = 0.3$, and $x_e = 1$. Moreover, we have fixed the $SU(5)$ gauge symmetry-breaking scale $M_{\alpha}$ to $M_{\text{GUT}} = 2 \times 10^{16}$ GeV and $n_s$ to the central value (0.9655) of the bound from Planck's data. Finally, the soft masses have been fixed at $12.5$ TeV, in order to avoid the dimension-5 proton decay. We restrict ourselves to the parameter region with the largest possible values of $r$ observable by the near-future experiments highlighted below. 

In our analysis the soft SUSY contributions to the inflationary trajectory are suppressed, while the radiative and SUGRA corrections, parametrised by $\gamma$ and $\sigma$, play the dominant role. To keep the SUGRA expansion under control we impose $S_0 \leq m_P$. We further require $\sigma \gtrsim -1$ and $\{2 \times 10^6 \lesssim T_r \lesssim 2 \times 10^9\}$ GeV. These constraints appear in Figs. \ref{fig:results1} and \ref{fig:results2} as the boundaries of the allowed region in the $\tilde{\gamma}-\sigma$ plane.
 
\subsection{Observable Primordial Gravitational Waves} \label{sec6_2}
 \begin{figure}[t]
	\centering \includegraphics[width=7.85cm]{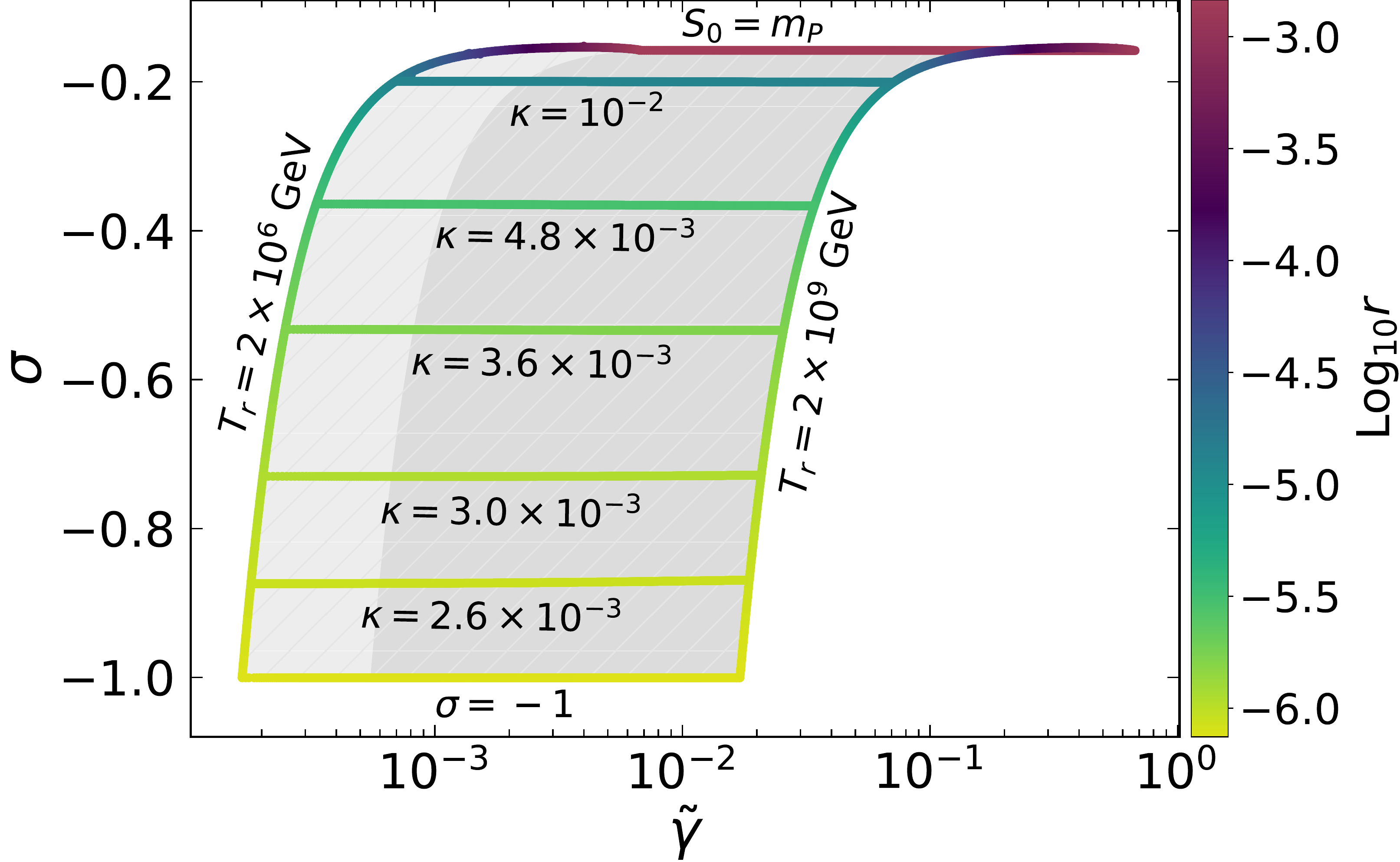}
	\centering \includegraphics[width=7.85cm]{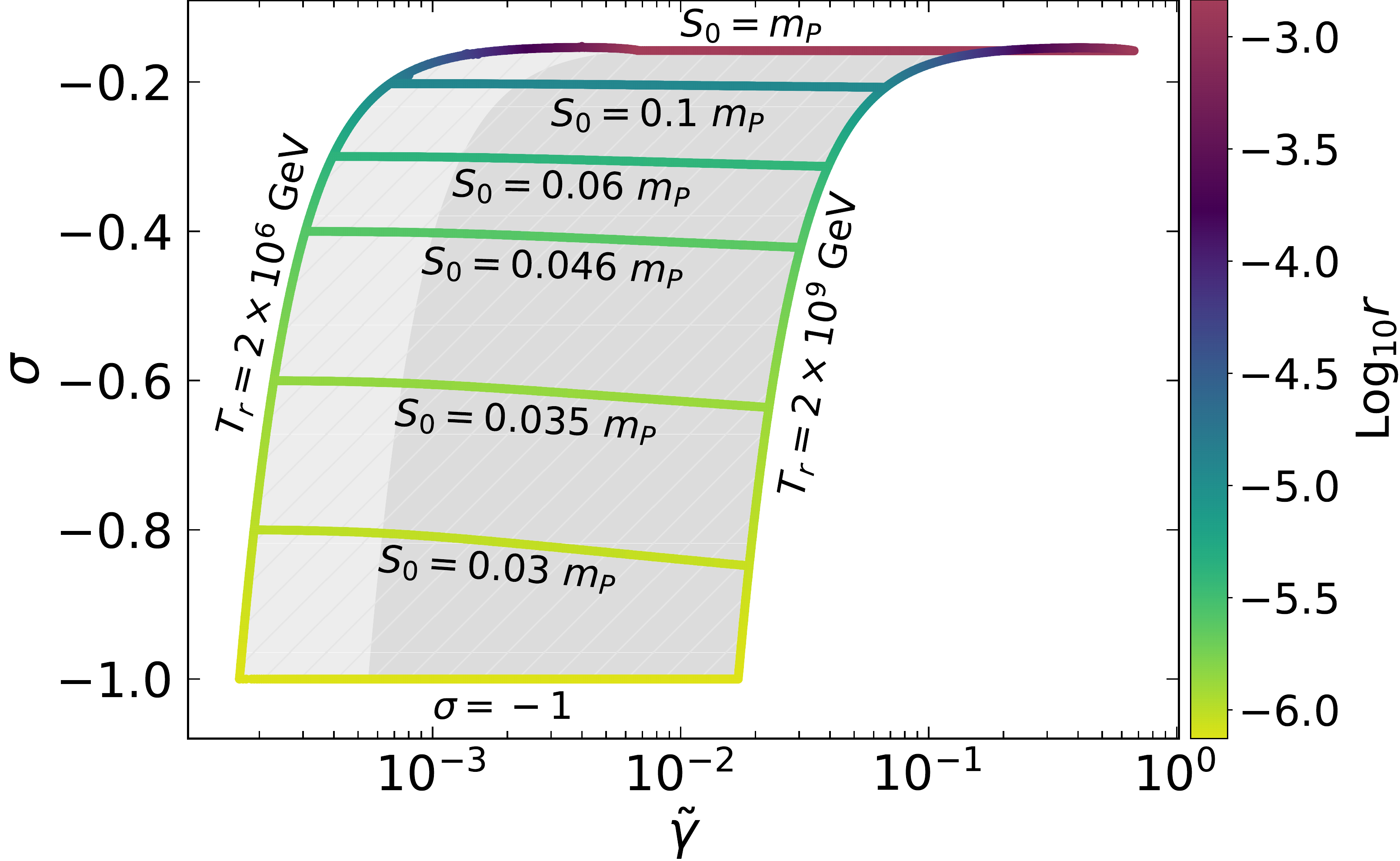}
	\centering \includegraphics[width=7.85cm]{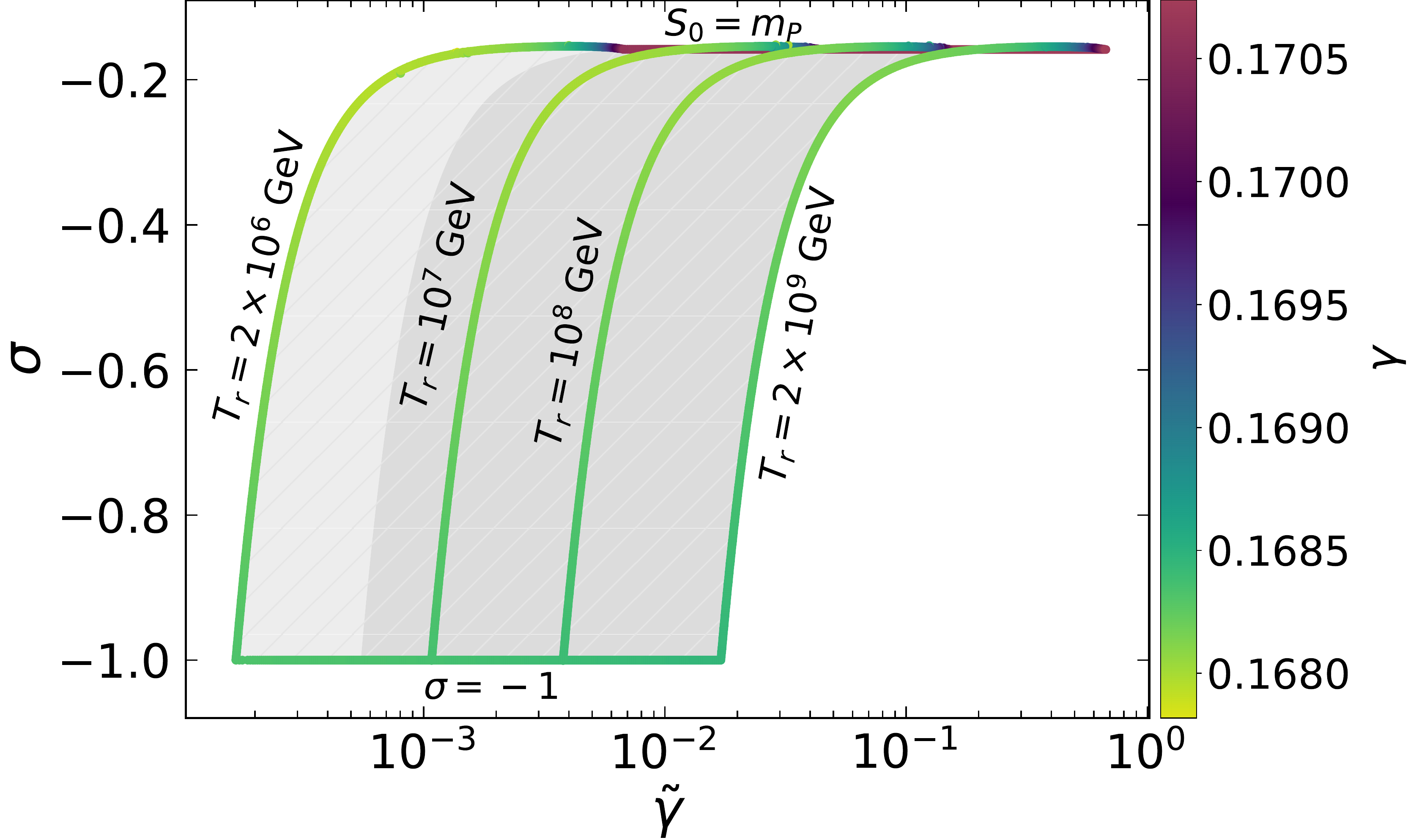}
	\caption{Variation of $\kappa$, $S_0$ and reheat temperature $T_r$ in $\tilde{\gamma}-\sigma$ plane. The color map displays the range of tensor to scalar ratio $r$ in the top two panels and $\gamma$ in the bottom panel. The light shaded region corresponds to the neutrino dominant channel whereas, the dark shaded region corresponds to Higgsino dominant channel.}
	\label{fig:results1}
\end{figure} 

The tensor-to-scalar ratio $r$ is the canonical measure of primordial gravitational waves and the next-generation experiments are gearing up to measure it. One of the highlights of PRISM \cite{Andre:2013afa} is to detect $r$ as low as $5 \times 10^{-4}$, and a major goal of LiteBIRD \cite{Matsumura:2013aja} is to attain a measurement of $r$ within an uncertainty of $\delta r = 0.001$. Furthermore, the CORE \cite{Finelli:2016cyd} experiment is forecast to have sensitivity to $r$ as low as $10^{-3}$, and PIXIE \cite{Kogut:2011xw} aims to measure $r < 10^{-3}$ at the 5$\sigma$ level. Other future missions include CMB-S4 \cite{Abazajian:2019eic}, which has the goal of detecting $r \gtrsim 0.003$ at greater than 5$\sigma$ or, in the absence of a detection, reaching an upper limit of $r < 0.001$ at the 95$\%$ confidence level, and PICO \cite{SimonsObservatory:2018koc}, which aims to detect $r = 5 \times 10^{-4}$ at 5$\sigma$.

The explicit dependence of $r$ on $\kappa$ and the $SU(5)$ symmetry-breaking scale $M_{\alpha}$ is given by the following approximate relation obtained by using the normalization constraint on $\Delta_{R}$
\begin{equation} \label{eq:rkappaexplicit}
	r \simeq \left( \frac{2 \kappa^2}{3 \pi^2 \Delta_{R}^2} \right) \left(\frac{M_{\alpha}}{m_P}\right)^4\,.
\end{equation}
One can see that, for fixed $M_{\alpha}$, large $r$ values should be obtained with large $\kappa$, which is indeed the case. For $M_{\alpha} \simeq 2 \times 10^{16}$ GeV, as used in our analysis, it can readily be checked that the above equation gives $r \sim 0.0014$ for $\kappa \sim 0.1$, and $r \sim 6 \times 10^{-7}$ for $\kappa \sim 0.002$. These approximate values are very close to the actual values obtained in our numerical calculations, and the  larger tensor modes can be detected by future experiments.
\begin{figure}[t]
	\centering \includegraphics[width=7.85cm]{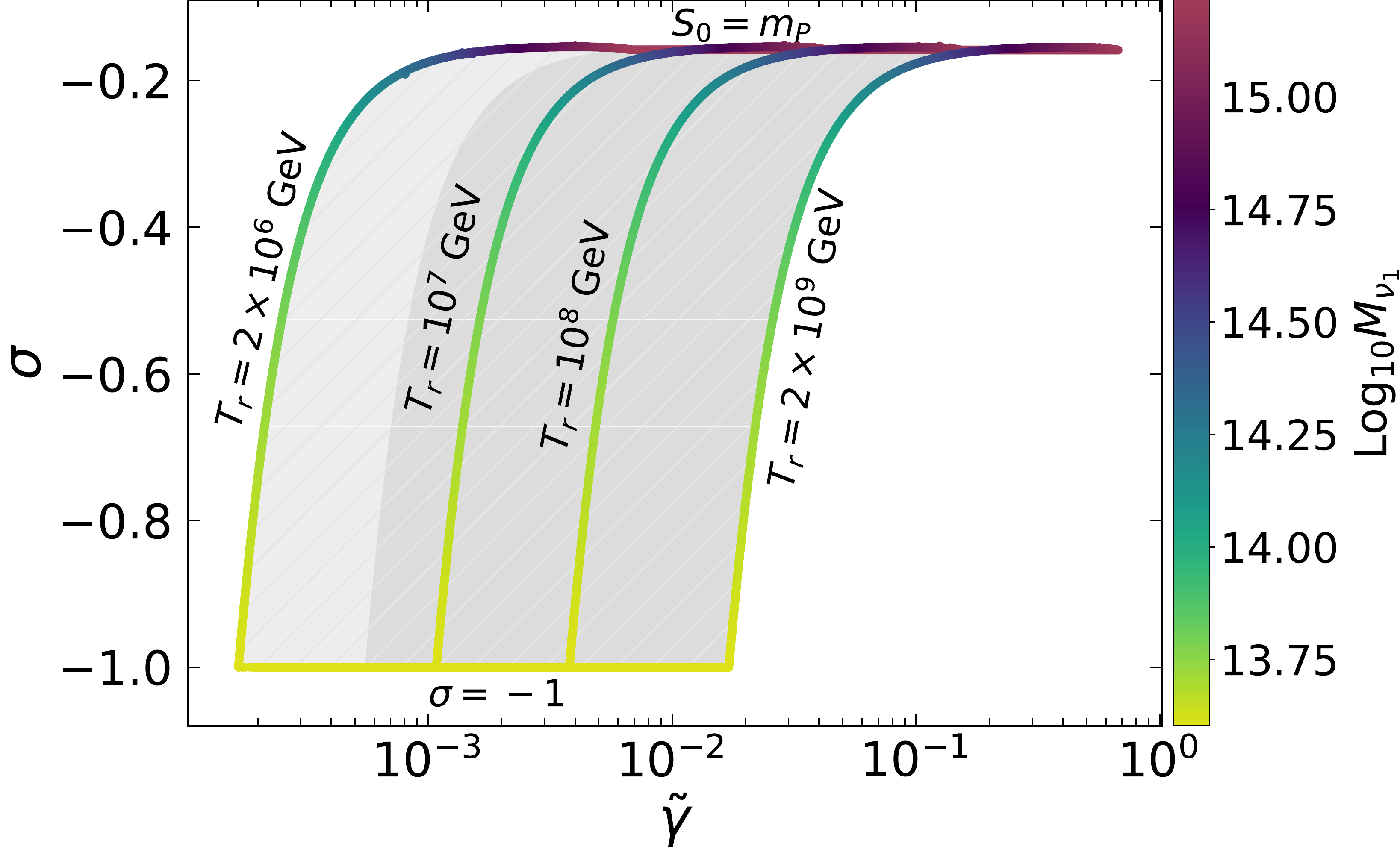}
	\centering \includegraphics[width=7.85cm]{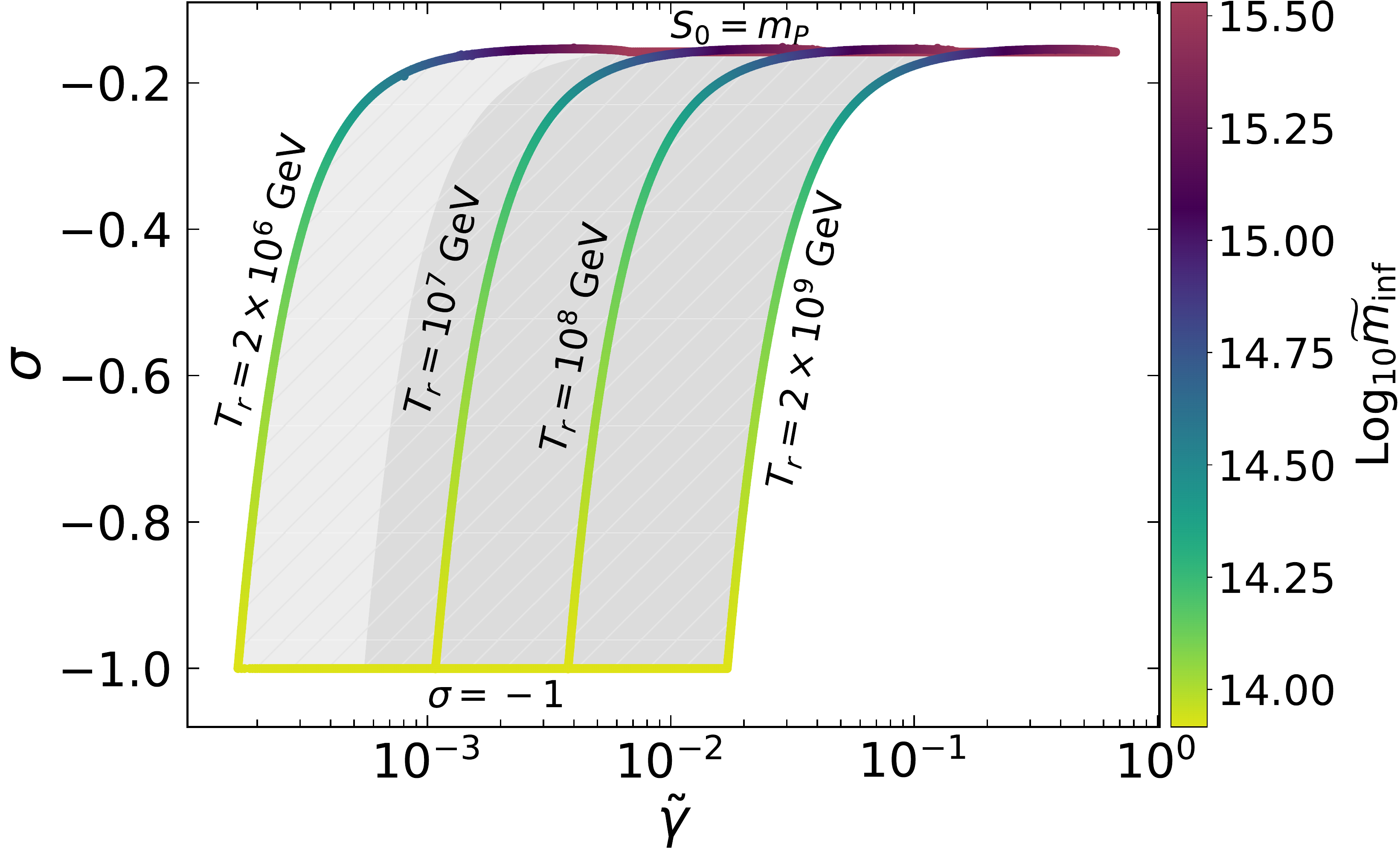}
	\caption{Variation of reheat temperature $T_r$ in $\tilde{\gamma}-\sigma$ plane. The color map displays the range of right handed neutrino mass $M_1$ in the left panel and inflaton mass $m_{\text{inf}}$ in the right panel. The light shaded region corresponds to the neutrino dominant channel whereas the dark shaded region corresponds to Higgsino dominant channel.}
	\label{fig:results2}
\end{figure}

In the leading order slow-roll approximation, the spectral index $n_s$ and tensor to scalar ratio $r$ are given by
\begin{eqnarray}
	n_s &\simeq& 1 + \left( \frac{m_P}{M_{\alpha}} \right)^2 \Bigg[ -\frac{z^{'}(x_0)}{z(x_0)^3} \bigg( 4 \left( \frac{M_{\alpha}}{m_P} \right)^2 x_0 \left( 2 \gamma - \frac{1}{3} \right) \nonumber \\
	&+& 4 \left( \frac{M_{\alpha}}{m_P} \right)^4 x_0^3 \left( \frac{14}{3} \gamma + 16 \gamma^2  - \frac{5}{9} + \frac{9}{2} \sigma \right) + \frac{275 \kappa^2}{16 \pi^2} F^{'}(5 x_0)  \bigg) \nonumber \\
	&+&  \frac{1}{z(x_0)^2} \bigg( 4 \left( \frac{M_{\alpha}}{m_P} \right)^2 \left( 2 \gamma - \frac{1}{3} \right) + 12 \left( \frac{M_{\alpha}}{m_P} \right)^4 x_0^2 \left( \frac{14}{3} \gamma + 16 \gamma^2  - \frac{5}{9} + \frac{9}{2} \sigma \right) \nonumber \\ 
	&+& \frac{275\kappa^2}{16 \pi^2} F^{''}(5 x_0)  \bigg) \Bigg] \Bigg[ 1 + \left( \frac{m_P}{M_{\alpha}} \right)^2 x_0^2 \left( 2 \gamma - \frac{1}{3} \right) + \frac{275 \kappa^2}{16 \pi^2} F(5 x_0) \Bigg]^{-1}\,,
\end{eqnarray}
\begin{eqnarray}
\text{and}\quad	r &\simeq& 3 \left( \frac{m_P}{M_{\alpha}} \right)^2 \frac{1}{z(x_0)^2} \Bigg[ 4 \left( \frac{M_{\alpha}}{m_P} \right)^4 x_0 \left( 2 \gamma - \frac{1}{3} \right) \nonumber \\
	&+& 4 \left( \frac{M_{\alpha}}{m_P} \right)^4 x_0^3 \left( \frac{14}{3} \gamma + 16 \gamma^2  - \frac{5}{9} + \frac{9}{2} \sigma \right) 
	+ \frac{275 \kappa^2}{16 \pi^2} F^{'}(5 x_0)  \Bigg]\,,
\end{eqnarray}
with 
\begin{equation*}
	z(x_0) \simeq \sqrt{1 + \left( \frac{M_{\alpha}}{m_P} \right)^2 x_0^2} \,.
\end{equation*}
Solving these two equations simultaneously, we obtain $\gamma \sim 0.17027$, $\sigma \sim -0.15726$ for $S_0 \simeq m_P$, $n_s \simeq 0.9655$, $r \simeq 0.0013$. Similarly, for $S_0 \simeq 0.027 m_P$, $n_s \simeq 0.9655$, $r \simeq 7.5 \times 10^{-7}$ we obtain $\gamma \sim 0.1683$, $\sigma \sim -1$. Both these estimates are in good agreement with our numerical results displayed in Figs. \ref{fig:results1} and \ref{fig:results2}. Thus, for couplings $0.1678 \lesssim \gamma \lesssim 0.1707$ and $-1 \lesssim \sigma \lesssim -0.1525$, we obtain $n_s$ compatible with the Planck constraints and $r$ in the $7.5 \times 10^{-7} - 1.5 \times 10^{-3}$ range. At the same time, given our chosen values of $M_\alpha$ and $M_{\text{SUSY}}$ that yield these results, the proton lifetime lies above the lower bound from the Super-K experiment and should be testable by the future Hyper-K experiment.

Fig. \ref{fig:results2} is of particular importance in connection with the non-thermal leptogenesis and shows the variation of $T_r$ in the $\tilde{\gamma}-\sigma$ plane. The color map displays the range of right handed neutrino mass $(4.0 \times 10^{13} \lesssim M_{\nu_1} \lesssim 1.6 \times 10^{15})$ GeV in the left panel and inflaton mass $(8.3 \times 10^{13} \lesssim \widetilde{m}_{\text{inf}} \lesssim 3.3 \times 10^{15})$ GeV in the right panel. Imposing the kinematic condition,
\begin{equation}
	\frac{\widetilde{m}_{\text{inf}}}{M_{\nu_1}} \geq 2\,,
\end{equation}
and using $n_L/s = 3 \times 10^{-10}$, Eq. \eqref{fin1} can be written as,
\begin{equation}
	\frac{\Gamma_{\text{inf}\rightarrow \nu^{c} \nu^{c}}}{\Gamma_{\text{inf}\rightarrow \bar{H} H}} = \left(\frac{T_r}{2 \times 10^6 \text{ GeV}}\right)\,, \label{nls}
\end{equation}
which implies $T_r > 4 \times 10^6$ GeV for the Higgsino-dominant decay of the inflaton, and $T_r < 4 \times 10^6$ GeV for the neutrino-dominant channel.

\subsection{BBN Constraints on Reheating Temperature and Gravitino Cosmology} \label{sec6_3}

Another important constraint on $T_{r}$ comes from gravitino cosmology, as it depends on the SUSY-breaking mechanism and the gravitino mass. As noted in \cite{Ahmed:2021dvo,Okada:2015vka,Rehman:2017gkm,Ahmed:2022rwy}, one may consider the case of 
\\ $a)$ a stable gravitino as the LSP; 
\\ $b)$ an unstable long-lived  ($\tilde{\tau}\gtrsim 1$\,sec) gravitino with $m_{3/2} < 25$ TeV;  
\\ $c)$ an unstable short-lived ($\tilde{\tau} < 1$\,sec) gravitino with $m_{3/2} > 25$ TeV.\\

In models based on SUGRA, the relic abundance of a stable gravitino LSP is given \cite{Bolz:2000fu, Pradler:2006qh,Eberl:2020fml} by \footnote{Taking into account only the dominant QCD contributions to the gravitino production rate. In principle there are extra contributions descending from the EW sector, as mentioned in \cite{Pradler:2006qh} and recently revised in \cite{Eberl:2020fml}.}
\begin{equation}\label{omega}
	\Omega_{3/2} h^{2}=0.08\left(\frac{T_{r}}{10^{10} \; \text{GeV} }\right)\left(\frac{m_{3/2}}{1 \; \text{TeV}}\right)\left(1+\frac{m_{\tilde{g}}^2}{3m_{3/2}^{2}}\right)~,
\end{equation}
\begin{figure}[t]
	\centering \includegraphics[width=9cm]{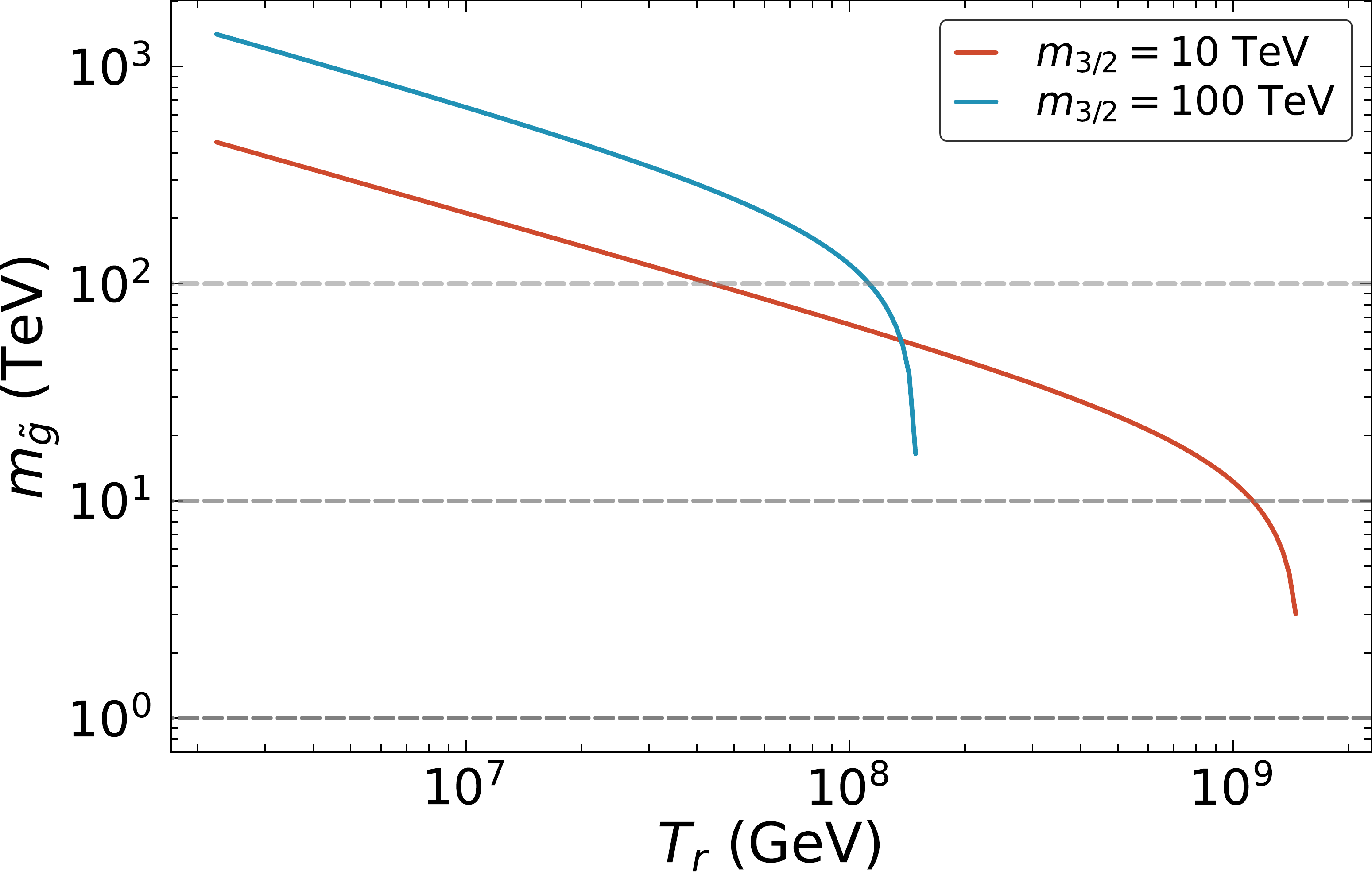}
	\caption{Possibility of the gravitino as the LSP for different values of its mass, $m_{3/2} = (10, 100)$ TeV, and the corresponding gluino masses.}
	\label{fig:reheat_gluino}
\end{figure} 
where $m_{\tilde{g}}$ is the gluino mass, $h$ is the present Hubble parameter in units of $100$ km $\text{sec}^{-1}\, \text{Mpc}^{-1}$, and $\Omega_{3/2} = \rho_{3/2}/\rho_c$\footnote{$\rho_{3/2}$ and $\rho_c$ are the gravitino energy density and the critical energy density of the present-day Universe, respectively.}. A stable  LSP  gravitino requires $m_{\tilde{g}}>m_{3/2}$, while current LHC  bounds on the gluino mass are around $2.3$ TeV \cite{Vami:2019slp}. It follows from Eq. \eqref{omega} that the overclosure limit, $\Omega_{3/2}h^2 < 1$, puts a severe upper bound on $T_{r}$, depending on $m_{3/2}$. Fig. \ref{fig:reheat_gluino} shows the gluino mass as a function of $T_r$ for $m_{3/2} = (10,\,100)$ TeV, with the condition that $\Omega_{3/2}h^2$ does not exceed the observed DM relic abundance, $\Omega h^2\leq 0.126$ \cite{planck18}, of the Universe. We see that $(m_{\tilde{g}}>m_{3/2})$ is satisfied for the entire range of the plotted parameter space, implying that the gravitino is consistently realized as the LSP in the model, and acts as a viable DM candidate.

When the gravitino is the next-to-LSP instead, the role of the LSP (and hence the DM) can be played by the lightest neutralino, $\tilde{\chi}_{1}^{0}$, which has two origins: thermal and non-thermal relic. Its thermal production consists of the standard freeze-out mechanism of the weakly interacting massive particles (WIMPs), whereas the non-thermal production proceeds via the decay of the gravitino, itself produced during the reheating process \cite{Kawasaki:2004qu,Kawasaki:2017bqm}. However, since the density of the thermal relic is strongly model-dependent, we do not take into account its effect in calculating the density parameter here. 

An unstable gravitino can be long-lived or short-lived \cite{Khlopov:1984pf,Ellis:1984eq}. For $m_{3/2}<25$ TeV, a gravitino lifetime of $\tilde{\tau}\gtrsim 1 \; \text{sec}$ \cite{Lazarides:2020zof,Kawasaki:2008qe} can be sufficiently long to result in the cosmological gravitino problem \cite{Khlopov:1984pf}. The fast decay of gravitino may affect the abundances of the light nuclei, thereby ruining the success of the big-bang nucelosynthesis (BBN) theory. To avoid this problem, one has to take into account the BBN bounds on $T_r$, which are conditioned on $m_{3/2}$ in gravity mediated SUSY-breaking as
\begin{equation} \label{eqlong}
\begin{split}
T_{r}&\lesssim 2\times 10^9  \;{\rm GeV } \quad {\rm for} \quad m_{3/2} \gtrsim 10 \; {\rm TeV}\,.
\end{split}
\end{equation}  
A long-lived gravitino scenario is therefore consistent with the BBN bounds \eqref{eqlong} for the entire $(12.5 \leq m_{3/2} \leq 25)$ TeV range, given the $T_r$ in our model. 

For a short-lived gravitino, the above $T_r$ bounds from BBN do not apply, and it can decay into the $\tilde{\chi}_{1}^{0}$ LSP, with the resultant abundance given by
 \begin{equation}\label{eq:a}
 \Omega_{\tilde{\chi}_{1}^{0}}h^{2}\simeq 2.8\times10^{11}\times Y_{3/2}\left(\frac{m_{\tilde{\chi}_{1}^{0}}}{1 \; \text{TeV}}\right)\,,
 \end{equation}
where the gravitino yield is defined as
 \begin{equation}\label{eq:b}
 Y_{3/2}\simeq2.3\times10^{-12}\left(\frac{T_{r}}{10^{10} \; \text{GeV}}\right)\,.
 \end{equation}
Requiring $\Omega_{\tilde{\chi}_{1}^{0}}h^{2} \lesssim 0.126$ leads to the relation
 \begin{eqnarray}\label{eqc}
 m_{\tilde{\chi}_{1}^{0}} \gtrsim 19.6\left(\frac{10^{11} \; \text{GeV}}{T_{r}}\right)~.
 \end{eqnarray}
The prediction of $\{2\times10^6 \lesssim T_{r}\lesssim 2\times10^9\}$ GeV by our model with gravity-mediated SUSY-breaking easily satisfies the $m_{\tilde{\chi}_{1}^{0}}\geq18$ GeV \cite{Hooper:2002nq} limit, thus making the short-lived gravitino a viable scenario also.

\section{Summary} \label{sec7}
To summarize, we have investigated various cosmological implications of a generic model based on the $SU(5)$ gauge symmetry formulated in the framework of no-scale supergravity, highlighting the issues of dimension-5 proton decay, inflation, gravitino, as well as baryogenesis via non-thermal leptogenesis. The breaking of $SU(5)$ gauge symmetry suffers from the magnetic monopole problem. Employing the shifted extension of hybrid inflation the monopole density is diluted and remain within the observable limit. The model yields large values of the tensor-to-scalar ratio ($r \sim 0.0015$) that are potentially measurable by future experiments and favors values of the scalar tilt $n_s$ that are consistent with current constraints. Moreover, the model avoids rapid proton decay via dimension-5 operators and also provides for baryogenesis via non-thermal leptogenesis, with low reheating temperature ($2 \times 10^6 \lesssim T_r \lesssim 2 \times 10^9$) GeV consistent with gravitino cosmology. 
 
\section*{Acknowledgements}
The authors are thankful to George K. Leonatris and Mansoor Ur Rehman for helpful
discussions. WA is partially supported by the Program for Excellent Talents in Hubei
Polytechnic University (21xjz22R, 21xjz21R, 21xjz20R) and SM would like to acknowledge
support from the ICTP through the Associates Programme (2022-2027).  



\begin{thebibliography}{99}

 \bibitem{reviews}
 K.~A.~Olive,
 Phys.\ Rept.\  {\bf 190} (1990) 307;
 A. D. Linde, {\it Particle  
 	Physics and
 	Inflationary Cosmology} (Harwood, Chur, Switzerland, 1990); 
 D.~H.~Lyth and A.~Riotto,
 {\it Phys.\ Rep.}  {\bf 314} (1999) 1
 [arXiv:hep-ph/9807278];
 J.~Martin, C.~Ringeval and V.~Vennin,
 Phys.\ Dark Univ.\  {\bf 5-6}, 75-235 (2014)
 [arXiv:1303.3787 [astro-ph.CO]];
 J.~Martin, C.~Ringeval, R.~Trotta and V.~Vennin,
 JCAP {\bf 1403} (2014) 039
 [arXiv:1312.3529 [astro-ph.CO]];
 J.~Martin,
 Astrophys.\ Space Sci.\ Proc.\  {\bf 45}, 41 (2016)
 [arXiv:1502.05733 [astro-ph.CO]].
 \bibitem{planck18}
 Y.~Akrami {\it et al.} [Planck Collaboration],
 arXiv:1807.06211 [astro-ph.CO].
 \bibitem{no-scale}
 E.~Cremmer, S.~Ferrara, C.~Kounnas and D.~V.~Nanopoulos,
 Phys.\ Lett.\ B {\bf 133} (1983) 61.
 \bibitem{Ellis:1983sf} 
 J.~R.~Ellis, A.~B.~Lahanas, D.~V.~Nanopoulos and K.~Tamvakis,
 Phys.\ Lett.\  {\bf 134B}, 429 (1984).
 \bibitem{LN}
 A.~B.~Lahanas and D.~V.~Nanopoulos,
 Phys.\ Rept.\  {\bf 145} (1987) 1.
 \bibitem{Witten}
 E.~Witten,
 Phys.\ Lett.\  {\bf 155B} (1985) 151.
 \bibitem{ENO8}
 J.~Ellis, D.~V.~Nanopoulos and K.~A.~Olive,
 Phys.\ Rev.\ D {\bf 89} (2014) 4,  043502
 [arXiv:1310.4770 [hep-ph]];
 J.~Ellis, M.~A.~G.~Garc{\' i}a, D.~V.~Nanopoulos and K.~A.~Olive,
 JCAP {\bf 1510}, 003 (2015)
 [arXiv:1503.08867 [hep-ph]].
 S.~F.~King and E.~Perdomo,
 JHEP {\bf 1905}, 211 (2019)
 [arXiv:1903.08448 [hep-ph]].
 J.~Ellis, M.~A.~G.~Garc{\' i}a, N.~Nagata, D.~V.~Nanopoulos and K.~A.~Olive,
 JCAP {\bf 1611}, no. 11, 018 (2016)
 [arXiv:1609.05849 [hep-ph]].
 \bibitem{EGNNO45}
 J.~Ellis, M.~A.~G.~Garcia, N.~Nagata, D.~V.~Nanopoulos and K.~A.~Olive,
 Phys.\ Lett.\ B {\bf 797}, 134864 (2019)
 [arXiv:1906.08483 [hep-ph]];
 J.~Ellis, M.~A.~G.~Garcia, N.~Nagata, D.~V.~Nanopoulos and K.~A.~Olive,
 JCAP {\bf 2001}, no. 01, 035 (2020)
 [arXiv:1910.11755 [hep-ph]].
 E.~Dudas, T.~Gherghetta, Y.~Mambrini and K.~A.~Olive,
 Phys.\ Rev.\ D {\bf 96}, no. 11, 115032 (2017)
 [arXiv:1710.07341 [hep-ph]];
 K.~Kaneta, Y.~Mambrini, K.~A.~Olive and S.~Verner,
 Phys.\ Rev.\ D {\bf 101}, no. 1, 015002 (2020)
 [arXiv:1911.02463 [hep-ph]].
 \bibitem{Starobinsky:1980te}
 A.~A.~Starobinsky,
 Phys. Lett. B \textbf{91}, 99-102 (1980)
 \bibitem{ENO}
 J.~Ellis, D.~V.~Nanopoulos and K.~A.~Olive,
 Phys.\ Rev.\ Lett.\  {\bf 111}, 111301 (2013)
 [Erratum-ibid.\  {\bf 111}, no. 12, 129902 (2013)]
 [arXiv:1305.1247 [hep-th]].
 \bibitem{CFKN}
 E.~Cremmer, S. ~Ferrara, C. ~Kounnas and D. V. ~Nanopoulos, Phys. Lett. B{\bf133}, 61 (1983).
 \bibitem{EKN}
 J. R. Ellis, C. Kounnas and D. V. Nanopoulos, Nucl. Phys.B {\bf247}, 373 (1984).

 \bibitem{Einhorn:2009bh}
 M.~B.~Einhorn and D.~R.~T.~Jones,
 JHEP {\bf 1003}, 026 (2010)
 [arXiv:0912.2718 [hep-ph]].
 \bibitem{Ferrara:2010in}
 S.~Ferrara, R.~Kallosh, A.~Linde, A.~Marrani and A.~Van Proeyen,
 Phys.\ Rev.\ D {\bf 83}, 025008 (2011)
 [arXiv:1008.2942 [hep-th]].
 \bibitem{Ferrara:2010yw}
 S.~Ferrara, R.~Kallosh, A.~Linde, A.~Marrani and A.~Van Proeyen,
 Phys.\ Rev.\ D {\bf 82}, 045003 (2010)
 [arXiv:1004.0712 [hep-th]].
 \bibitem{Ellis:2020lnc}
 J.~Ellis, M.~A.~G.~Garcia, N.~Nagata, N.~D.~V., K.~A.~Olive and S.~Verner,
 Int. J. Mod. Phys. D \textbf{29} (2020) no.16, 2030011
 [arXiv:2009.01709 [hep-ph]].
 \bibitem{Ahmed:2018jlv}
 W.~Ahmed and A.~Karozas,
 Phys. Rev. D \textbf{98}, no.2, 023538 (2018)
 [arXiv:1804.04822 [hep-ph]].
 \bibitem{Ahmed:2021dvo}
 W.~Ahmed, A.~Karozas and G.~K.~Leontaris,
 Phys. Rev. D \textbf{104}, no.5, 055025 (2021)
 [arXiv:2104.04328 [hep-ph]].
 \bibitem{Dvali:1994ms}
 G.~R.~Dvali, Q.~Shafi and R.~K.~Schaefer,
 ``Large scale structure and supersymmetric inflation without fine tuning,''
 Phys. Rev. Lett. \textbf{73}, 1886-1889 (1994)
 [arXiv:hep-ph/9406319 [hep-ph]].
 \bibitem{Copeland:1994vg}
 E.~J.~Copeland, A.~R.~Liddle, D.~H.~Lyth, E.~D.~Stewart and D.~Wands,
 ``False vacuum inflation with Einstein gravity,''
 Phys. Rev. D \textbf{49}, 6410-6433 (1994)
 [arXiv:astro-ph/9401011 [astro-ph]].
 \bibitem{Linde:1997sj}
 A.~D.~Linde and A.~Riotto,
 ``Hybrid inflation in supergravity,''
 Phys. Rev. D \textbf{56}, R1841-R1844 (1997)
 [arXiv:hep-ph/9703209 [hep-ph]].
 \bibitem{Senoguz:2004vu}
 V.~N.~Senoguz and Q.~Shafi,
 ``Reheat temperature in supersymmetric hybrid inflation models,''
 Phys. Rev. D \textbf{71}, 043514 (2005)
 [arXiv:hep-ph/0412102 [hep-ph]].
 \bibitem{Rehman:2009nq}
 M.~U.~Rehman, Q.~Shafi and J.~R.~Wickman,
 ``Supersymmetric Hybrid Inflation Redux,''
 Phys. Lett. B \textbf{683}, 191-195 (2010)
 [arXiv:0908.3896 [hep-ph]].
 \bibitem{Buchmuller:2014epa}
 W.~Buchm\"uller, V.~Domcke, K.~Kamada and K.~Schmitz,
 ``Hybrid Inflation in the Complex Plane,''
 JCAP \textbf{07}, 054 (2014)
 [arXiv:1404.1832 [hep-ph]].
 
 \bibitem{Abid:2021jvn}
 M.~M.~A.~Abid, M.~Mehmood, M.~U.~Rehman and Q.~Shafi,
 JCAP \textbf{10}, 015 (2021)
 [arXiv:2107.05678 [hep-ph]].
 
 
 \bibitem{Rehman:2018nsn}
 M.~U.~Rehman, Q.~Shafi and U.~Zubair,
 Phys. Rev. D \textbf{97}, no.12, 123522 (2018)
 [arXiv:1804.02493 [hep-ph]].
 
  \bibitem{Kibble:1976sj}
  T.~W.~B.~Kibble,
  J. Phys. A \textbf{9}, 1387-1398 (1976)
 
 
 \bibitem{Khalil:2010cp} 
 S.~Khalil, M.~U.~Rehman, Q.~Shafi and E.~A.~Zaakouk,
 Phys.\ Rev.\ D {\bf 83}, 063522 (2011)
 [arXiv:1010.3657 [hep-ph]].
 
  
  \bibitem{Masoud:2019gxx}
  M.~A.~Masoud, M.~U.~Rehman and Q.~Shafi,
  JCAP \textbf{04}, 041 (2020)
  [arXiv:1910.07554 [hep-ph]].
 
 \bibitem{Rehman:2014rpa}
 M.~U.~Rehman and U.~Zubair,
 Phys. Rev. D \textbf{91}, 103523 (2015)
 [arXiv:1412.7619 [hep-ph]].
 
 \bibitem{Rehman:2012gd} 
 M.~U.~Rehman and Q.~Shafi,
 Phys.\ Rev.\ D {\bf 86}, 027301 (2012)
 [arXiv:1202.0011 [hep-ph]].
 
 \bibitem{Ahmed:2022vlc}
 W.~Ahmed, A.~Karozas, G.~K.~Leontaris and U.~Zubair,
 JCAP \textbf{06}, no.06, 027 (2022)
 
 

 
 
 \bibitem{Barr:2005xya}
 S.~M.~Barr, B.~Kyae and Q.~Shafi,
 [arXiv:hep-ph/0511097 [hep-ph]].
 
 \bibitem{Fallbacher:2011xg}
 M.~Fallbacher, M.~Ratz and P.~K.~S.~Vaudrevange,
 Phys. Lett. B \textbf{705}, 503-506 (2011)

 
 

 
 \bibitem{Nagata:2013ive}
 N.~Nagata,
 ``Proton Decay in High-scale Supersymmetry,''
 doi:10.15083/00006623
 
 
 \bibitem{Super-Kamiokande:2016exg}
 K.~Abe \textit{et al.} [Super-Kamiokande],
 Phys. Rev. D \textbf{95}, no.1, 012004 (2017)
 [arXiv:1610.03597 [hep-ex]].
 
 
 \bibitem{Hyper-Kamiokande:2018ofw}
 K.~Abe \textit{et al.} [Hyper-Kamiokande],
 [arXiv:1805.04163 [physics.ins-det]].
 
 
 \bibitem{Coleman-Weinberg: 1973}
 S.R.~Coleman and E.J.~Weinberg,
 Phys.\ Rev.\ D\  {\bf 7}, 1888 (1973)
 [arXiv:hep-ph/].
 
 
 \bibitem{Nilles:1983ge}
 H.~P.~Nilles,
 Phys. Rept. \textbf{110}, 1-162 (1984)
 
 
 
 
 \bibitem{Endo:2007sz}
 M.~Endo, F.~Takahashi and T.~T.~Yanagida,
 Phys. Rev. D \textbf{76}, 083509 (2007)
 [arXiv:0706.0986 [hep-ph]].
 
 \bibitem{Pallis:2011gr}
 C.~Pallis and N.~Toumbas,
 JCAP \textbf{12}, 002 (2011)
 [arXiv:1108.1771 [hep-ph]].
 
 \bibitem{Lazarides:1998qx}
 G.~Lazarides and N.~D.~Vlachos,
 Phys. Lett. B \textbf{441}, 46-51 (1998)
 [arXiv:hep-ph/9807253 [hep-ph]].




\bibitem{Liddle:2003as}
A.~R.~Liddle and S.~M.~Leach,
Phys. Rev. D \textbf{68}, 103503 (2003)
[arXiv:astro-ph/0305263 [astro-ph]].


\bibitem{Hamaguchi:2002vc}
K.~Hamaguchi,
[arXiv:hep-ph/0212305 [hep-ph]].




\bibitem{Planck:2018vyg}
N.~Aghanim \textit{et al.} [Planck],
Astron. Astrophys. \textbf{641}, A6 (2020)
[erratum: Astron. Astrophys. \textbf{652}, C4 (2021)]
[arXiv:1807.06209 [astro-ph.CO]].



\bibitem{Kuzmin:1985mm}
V.~A.~Kuzmin, V.~A.~Rubakov and M.~E.~Shaposhnikov,
Phys. Lett. B \textbf{155}, 36 (1985)

\bibitem{Fukugita:1986hr}
M.~Fukugita and T.~Yanagida,
Phys. Lett. B \textbf{174}, 45-47 (1986)

 
 \bibitem{Andre:2013afa}
 P.~Andre \textit{et al.} [PRISM],
 [arXiv:1306.2259 [astro-ph.CO]].
 
 
 
 \bibitem{Matsumura:2013aja}
 T.~Matsumura \textit{et al.} [Mission design of LiteBIRD],
 J. Low Temp. Phys. \textbf{176}, 733 (2014)
 [arXiv:1311.2847 [astro-ph.IM]].
 
  
  \bibitem{Finelli:2016cyd}
  F.~Finelli \textit{et al.} [CORE],
  JCAP \textbf{04}, 016 (2018)
  [arXiv:1612.08270 [astro-ph.CO]].
 
 
 \bibitem{Kogut:2011xw}
 A.~Kogut, D.~J.~Fixsen, D.~T.~Chuss, J.~Dotson, E.~Dwek, M.~Halpern, G.~F.~Hinshaw, S.~M.~Meyer, S.~H.~Moseley, M.~D.~Seiffert, D.~N.~Spergel and E.~J.~Wollack,
 JCAP \textbf{07}, 025 (2011)
 [arXiv:1105.2044 [astro-ph.CO]].
 
 

 
 
 
 \bibitem{Abazajian:2019eic}
 K.~Abazajian, G.~Addison, P.~Adshead, Z.~Ahmed, S.~W.~Allen, D.~Alonso, M.~Alvarez, A.~Anderson, K.~S.~Arnold and C.~Baccigalupi, \textit{et al.}
 [arXiv:1907.04473 [astro-ph.IM]].
 
 \bibitem{SimonsObservatory:2018koc}
 P.~Ade \textit{et al.} [Simons Observatory],
 JCAP \textbf{02}, 056 (2019)
 [arXiv:1808.07445 [astro-ph.CO]].
 


\bibitem{Okada:2015vka}
N.~Okada and Q.~Shafi,
Phys. Lett. B \textbf{775}, 348-351 (2017)
[arXiv:1506.01410 [hep-ph]].

  
\bibitem{Rehman:2017gkm}
M.~U.~Rehman, Q.~Shafi and F.~K.~Vardag,
Phys. Rev. D \textbf{96}, no.6, 063527 (2017)
[arXiv:1705.03693 [hep-ph]].

\bibitem{Ahmed:2022rwy}
W.~Ahmed, M.~Junaid, S.~Nasri and U.~Zubair,
Phys. Rev. D \textbf{105}, no.11, 115008 (2022)
[arXiv:2202.06216 [hep-ph]].

\bibitem{Bolz:2000fu}
M.~Bolz, A.~Brandenburg and W.~Buchmuller,
Nucl. Phys. B \textbf{606}, 518-544 (2001)
[erratum: Nucl. Phys. B \textbf{790}, 336-337 (2008)]
[arXiv:hep-ph/0012052 [hep-ph]].




\bibitem{Pradler:2006qh}
J.~Pradler and F.~D.~Steffen,
Phys. Rev. D \textbf{75}, 023509 (2007)
[arXiv:hep-ph/0608344 [hep-ph]].
022



\bibitem{Eberl:2020fml}
H.~Eberl, I.~D.~Gialamas and V.~C.~Spanos,
Phys. Rev. D \textbf{103}, no.7, 075025 (2021)
[arXiv:2010.14621 [hep-ph]].


\bibitem{Vami:2019slp}
T.~A.~Vami [ATLAS and CMS],
PoS \textbf{LHCP2019}, 168 (2019)
[arXiv:1909.11753 [hep-ex]].




\bibitem{Kawasaki:2004qu}
M.~Kawasaki, K.~Kohri and T.~Moroi,
Phys. Rev. D \textbf{71}, 083502 (2005)
[arXiv:astro-ph/0408426 [astro-ph]].

\bibitem{Kawasaki:2017bqm}
M.~Kawasaki, K.~Kohri, T.~Moroi and Y.~Takaesu,
Phys. Rev. D \textbf{97}, no.2, 023502 (2018)
[arXiv:1709.01211 [hep-ph]].

\bibitem{Khlopov:1984pf}
M.~Y.~Khlopov and A.~D.~Linde,
Phys. Lett. B \textbf{138}, 265-268 (1984)

\bibitem{Ellis:1984eq}
J.~R.~Ellis, J.~E.~Kim and D.~V.~Nanopoulos,
Phys. Lett. B \textbf{145}, 181-186 (1984)


\bibitem{Lazarides:2020zof}
G.~Lazarides, M.~U.~Rehman, Q.~Shafi and F.~K.~Vardag,
Phys. Rev. D \textbf{103}, no.3, 035033 (2021)
[arXiv:2007.01474 [hep-ph]].

\bibitem{Kawasaki:2008qe}
M.~Kawasaki, K.~Kohri, T.~Moroi and A.~Yotsuyanagi,
Phys. Rev. D \textbf{78}, 065011 (2008)
[arXiv:0804.3745 [hep-ph]].




\bibitem{Hooper:2002nq}
D.~Hooper and T.~Plehn,
Phys. Lett. B \textbf{562}, 18-27 (2003)
[arXiv:hep-ph/0212226 [hep-ph]].


\end{thebibliography}
\end{document}